\documentclass[journal,letterpaper,twoside,twocolumn]{IEEEtran}
\renewcommand{\IEEEQED}{\IEEEQEDopen} %
\usepackage{amsmath,amssymb,dsfont,graphicx,psfrag,cite}
\usepackage{color}
\usepackage[color,notcite,notref,final]{showkeys} %
\usepackage{balance} %
\definecolor{refkey}{rgb}{1,0.5,0} %
\definecolor{labelkey}{rgb}{1,0.5,0}
\usepackage{subfigure,multirow}

\newcommand{\eqlab}[2]{\begin{align} \label{#1} #2 \end{align}}
\newcommand{\eq}[1]{\begin{align*} #1 \end{align*}}
\newcommand{\SNR}[0]{\rho}
\newcommand{\MI}[0]{I}
\newcommand{\C}[0]{C}
\newcommand{\mf}[1]{\mathsf{#1}}
\newcommand{\set}[1]{\{#1\}}

\newcommand{\diag}{\mathop{\mathrm{diag}}}

\newcommand{\ld}{\ldots}
\newcommand{\inner}[2]{\left\langle#1,#2\right\rangle}
\newcommand{\ba}{\boldsymbol{a}}

\newcommand{\bb}{\boldsymbol{b}}

\newcommand{\bc}{\boldsymbol{c}}

\newcommand{\bh}{\boldsymbol{h}}
\newcommand{\bH}{\boldsymbol{H}}

\newcommand{\bn}{\boldsymbol{n}}

\newcommand{\bs}{\boldsymbol{s}}

\newcommand{\bx}{\boldsymbol{x}}
\newcommand{\bX}{{\mathop{\boldsymbol{X}}}}
\newcommand{\by}{\boldsymbol{y}}
\newcommand{\bY}{\boldsymbol{Y}}

\newcommand{\bZ}{\boldsymbol{Z}}
\newcommand{\bzero}{\boldsymbol{0}}

\newcommand{\bmu}{\boldsymbol{\mu}}

\newcommand{\mcB}{\mathcal{B}}
\newcommand{\mL}{{\mathbb{L}}}

\newcommand{\mN}{{\mathbb{N}}}
\newcommand{\mP}{{\mathbb{P}}}
\newcommand{\mS}{{\mathbb{S}}}
\newcommand{\mH}{{\mathbb{H}}}
\newcommand{\mG}{{\mathbb{G}}}
\newcommand{\mI}{{\mathbb{I}}}
\newcommand{\mD}{{\mathbb{D}}}
\newcommand{\mT}{{\mathbb{T}}}
\newcommand{\mX}{{\mathbb{X}}}
\newcommand{\mU}{{\mathbb{U}}}
\newcommand{\E}{{\mathds{E}}}
\newcommand{\R}{{\mathds{R}}}
\newcommand{\T}{{\mathrm{T}}}
\newcommand{\Es}{{E_\mathrm{s}}}
\newcommand{\Ebr}{{E_\mathrm{b}^\mathrm{r}}}

\newcommand{\Rc}{{R_\mathrm{c}}}
\newcommand{\tbs}{{\tilde{\boldsymbol{s}}}}
\newcommand{\tbx}{{\tilde{\boldsymbol{x}}}}
\newcommand{\tX}{{\tilde{\mX}}}
\newcommand{\ntbx}{{\mathring{\boldsymbol{x}}}}
\newcommand{\ntX}{{\mathring{\mX}}}

\newcommand{\sumi}{\sum_{i=0}^{M-1}}
\newcommand{\sumj}{\sum_{j=0}^{M-1}}
\newcommand{\sumk}{\sum_{k=0}^{m-1}}
\newcommand{\suml}{\sum_{l=0}^{M-1}}
\newcommand{\prodk}{\prod_{k=0}^{m-1}}
\newcommand{\pck}{P_{C_k}}
\newcommand{\tS}{\tilde{\mS}}

\newcommand{\overlay}[3]{\makebox[0mm][l]{\hspace*{#1}\raisebox{#2}[0ex][0ex]{#3}}}

\newtheorem{theorem}{Theorem}
\newtheorem{example}{Example}
\newtheorem{corollary}[theorem]{Corollary}
\newtheorem{definition}{Definition}
\newtheorem{lemma}[theorem]{Lemma}
\newtheorem{remark}{Remark}

\renewcommand{\markboth}[1]
  {\renewcommand{\leftmark}{#1}\renewcommand{\rightmark}{#1}}
\newcommand{\trash}[1]{}

\newlength{\xs}\newlength{\ys}

\renewcommand{\gamma}{g}

\title{Signal Shaping for BICM at Low SNR}
\author{Erik Agrell and Alex Alvarado
\thanks{Research supported by The British Academy and The Royal Society (via the Newton International Fellowship scheme), U.K., and by the European Community's Seventh's Framework Programme (FP7/2007-2013) under grant agreement No. 271986. This work was presented in part at the Information Theory and Applications (ITA) Workshop, San Diego, CA, February 2012, and at the IEEE International Symposium on Information Theory, Cambridge, MA, July 2012.

E.~Agrell is with the Dept.~of Signals and Systems, Chalmers Univ.~of Technology, SE-41296 G\"oteborg, Sweden (email: agrell@chalmers.se). A.~Alvarado is with the Dept.~of Engineering, University of Cambridge, Cambridge CB2 1PZ, United Kingdom (email: alex.alvarado@ieee.org).}}

\begin{document}
\maketitle

\begin{abstract}
The generalized mutual information (GMI) of bit-interleaved coded modulation (BICM) systems, sometimes called the BICM capacity, is investigated at low signal-to-noise ratio (SNR). The combinations of input alphabet, input distribution, and binary labeling that achieve the Shannon limit --1.59 dB are completely characterized. The main conclusion is that a BICM system with probabilistic shaping achieves the Shannon limit at low SNR if and only if it can be represented as a zero-mean linear projection of a hypercube. Hence, probabilistic shaping offers no extra degrees of freedom to optimize the low-SNR BICM-GMI, in addition to what is provided by geometrical shaping. The analytical conclusions are confirmed by numerical results, which also show that for a fixed input alphabet, probabilistic shaping can improve the BICM-GMI in the low and medium SNR range.
\end{abstract}


\section{Introduction}\label{Sec:Introduction}
The most important breakthrough for coded modulation (CM) in fading channels came in 1992, when Zehavi introduced the so-called bit-interleaved coded modulation (BICM) \cite{Zehavi92}, usually referred to as a pragmatic approach for CM \cite{Caire98,Fabregas08_Book}. Despite not being fully understood theoretically, BICM has been rapidly adopted in commercial systems such as wireless and wired broadband access networks, 3G/4G telephony, and digital video broadcasting, making it the de facto standard for current telecommunications systems \cite[Ch.~1]{Fabregas08_Book}.

Signal shaping refers to the use of non-equally spaced and/or non-equally likely symbols, i.e., \emph{geometrical shaping} and \emph{probabilistic shaping}, resp. Signal shaping has been studied during many years, cf.~\cite{Calderbank90,Fischer02_Book} and references therein. In the context of BICM, geometrical shaping was studied in \cite{Sommer00,LeGoff03,Barsoum07}, and probabilistic shaping, i.e., varying the probabilities of the bit streams, was first proposed in \cite{LeGoff04,Raphaeli04} and developed further in \cite{LeGoff05, LeGoff07, Fabregas10a,Peng12}. Probabilistic shaping offers another degree of freedom in the BICM design, which can be used to make the discrete input distribution more similar to the optimal distribution (which is in general unknown). This is particularly advantageous at low and medium SNR.

For the additive white Gaussian noise (AWGN) channel, the so-called Shannon Limit (SL) $-1.59~\text{dB}$ represents the average bit energy-to-noise ratio needed to transmit information reliably when the signal-to-noise ratio (SNR) tends to zero \cite{Verdu02,Prelov04}, i.e., in the wideband regime. 
When discrete input alphabets are considered at the transmitter and a BICM decoder is used at the receiver, the SL is not always achieved as first noticed in \cite{Martinez08b}. This was later shown to be caused by the selection of the binary labeling \cite{Stierstorfer09a}. The behavior of BICM in the wideband regime was studied in \cite{Martinez08b, Alvarado10c, Stierstorfer08a, Stierstorfer09a,Agrell10b} as a function of the alphabet ($\mX$) and the binary labeling ($\mL$), \emph{assuming a uniform input distribution}. First-order optimal (FOO) constellations were defined in \cite{Agrell10b} as the triplet $[\mX,\mP,\mL]$ that make a BICM system achieve the SL, where $\mP$ represents the input distribution.

In this paper, the results of \cite{Agrell10b} are generalized to nonuniform input distributions and give a complete characterization of FOO constellations for BICM in terms of $[\mX,\mP,\mL]$. More particularly, the geometrical and/or probabilistic shaping rules that should be applied to a constellation to make it FOO are found. The main conclusion is that probabilistic shaping offers no extra degrees of freedom in addition to what is provided by geometrical shaping for BICM in the wideband regime.%

\begin{figure*}
\newcommand{\scale}{0.85}
\psfrag{b1}[Br][Br][\scale]{information bits}
\psfrag{c1}[Bl][Bl][\scale]{$C_{0}$}
\psfrag{cm}[Bl][Bl][\scale]{$C_{m-1}$}
\psfrag{ENC}[cc][cc][\scale]{encoder(s)}
\psfrag{+}[cc][cc][\scale]{+}
\psfrag{INT}[cc][cc][\scale]{interleaver(s)}
\psfrag{bicmenc}[Bl][Bl][\scale]{BICM encoder}
\psfrag{bicmdec}[Br][Br][\scale]{BICM decoder}
\psfrag{ddd}[cc][cc][0.7][90]{$\cdots$}
\psfrag{MAP}[cc][cc][\scale]{$\Phi$}
\psfrag{x}[Bl][Bl][\scale]{$\bX$}
\psfrag{z}[Bl][Bl][\scale]{$\bZ$}
\psfrag{h}[Bl][Bl][\scale]{$\bH$}
\psfrag{y}[br][Br][\scale]{$\bY$}
\psfrag{DMAP}[cc][cc][\scale]{$\Phi^{-1}$}
\psfrag{l1}[Bl][Bl][\scale]{$L_0$}
\psfrag{lm}[Bl][Bl][\scale]{$L_{m-1}$}
\psfrag{DENC}[cc][cc][\scale]{decoder(s)}
\psfrag{DINT}[cc][cc][\scale]{deinterleaver(s)}
\psfrag{b1h}[Bl][Bl][\scale]{information bits}
\begin{center}
   \includegraphics[width=0.75\textwidth]{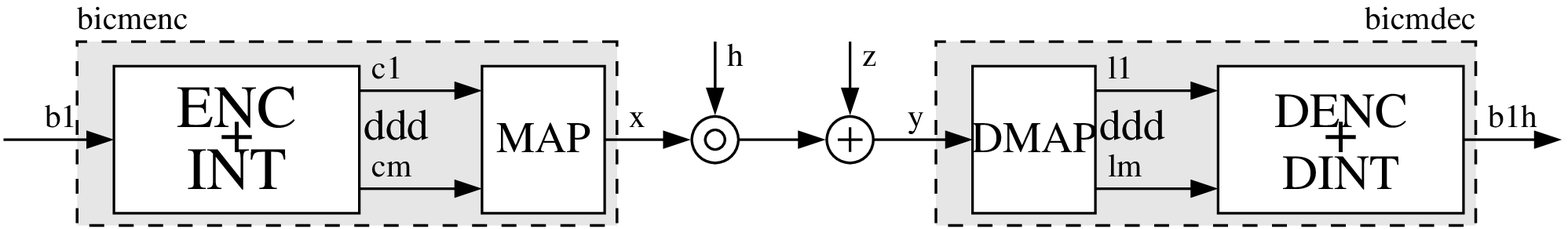}
      \caption{A generic BICM system, consisting of a BICM transmitter, the channel, and a BICM receiver.}
   \label{model}
\end{center}
\end{figure*}

\section{Preliminaries}\label{Sec:preliminaries}

\subsection{Notation}\label{Sec:preliminaries:Notation}

Bold italic letters $\bx$ denote row vectors. Block letters $\mX$ denote matrices or sometimes column vectors. The identity matrix is $\mI$. The inner product between two row vectors $\ba$ and $\bb$ is denoted by $\inner{\ba}{\bb}$ and their element-wise product by $\ba\circ\bb$. The Euclidean norm of the vector $\ba$ is denoted by $\|\ba\|$. Random variables are denoted by capital letters $X$ and random vectors by boldface capital vectors $\bX$. The probability density function (pdf) of the random vector $\bY$ is denoted by $p_{\bY}(\by)$ and the conditional pdf by $p_{\bY|\bX}(\by|\bx)$. A similar notation applies to probability mass functions of a random variable, which are denoted by $P_{\bY}(\by)$ and $P_{\bY|\bX}(\by|\bx)$. Expectations are denoted by $\E$.

The empty set is denoted by $\varnothing$ and the binary set by $\mcB\triangleq\set{0,1}$. The negation of a bit $b$ is denoted by $\bar{b} = 1-b$. Binary addition (exclusive-OR) of two bits $a$ and $b$ is denoted by $a \oplus b$. The same notation $a \oplus b$ denotes the integer that results from taking the bitwise exclusive-or of two integers $a$ and $b$.

\subsection{System Model}%

We consider transmissions over a discrete-time memoryless vectorial fast fading channel. The received vector at any discrete time instant is
\begin{align}\label{Fading_AWGN_channel}
\bY =\bH\circ\bX+\bZ
\end{align}
where $\bX$ is the channel input and $\bZ$ is Gaussian noise with zero mean and variance $N_0/2$ in each dimension \cite{Zehavi92}, \cite[App.~2.A]{Fabregas08_Book}. The channel is represented by the $N$-dimensional vector $\bH$. It contains the real fading coefficients $H_i$, which are random, possibly dependent, with the same pdf $p_H(h)$. We assume that $\bH$ and $N_0$ are perfectly known at the receiver or can be perfectly estimated, and that the technical requirements on $\bX$ and $\bH$ in \cite[Sec.~I-D]{Agrell10b} are satisfied.

The conditional transition pdf of the channel in \eqref{Fading_AWGN_channel} is
\begin{align}\label{ctpdf_AWGN}
p_{\bY|\bX,\bH}(\by|\bx,\bh) 	& = \frac{1}{(N_0\pi)^{N/2}}\exp{\biggl(-\frac{\|\by-\bh\circ\bx\|^2}{N_0}\biggr)}.
\end{align}
The SNR is defined as 
\begin{align}\label{snr.definition}
\SNR\triangleq \E[H^2]\frac{\Es}{N_0} = \Rc\frac{\Ebr}{N_0}
\end{align}
where $\Es \triangleq \E[\|\bX\|^2]$ is the average \emph{transmitted} symbol energy, $\Rc$ is the transmission rate in information bits per symbol, and $\Ebr \triangleq \E[H^2] \Es/\Rc$ is the average \emph{received} energy per information bit.

The generic BICM scheme in Fig.~\ref{model} is considered. The transmitter is, in the simplest case, a single binary encoder concatenated with an interleaver and a memoryless mapper $\Phi$. Multiple encoders and/or interleavers may be needed to achieve probabilistic shaping 
\cite{Raphaeli04, LeGoff05, LeGoff07, Fabregas10a}. At the receiver, using the channel output $\bY$, the demapper $\Phi^{-1}$ computes metrics $L_k$ for the individual coded bits $C_k$ with $k=0,\ld,m-1$, usually in the form of logarithmic likelihood ratios. These metrics are then passed to the deinterleaver(s) and decoder(s) to obtain an estimate of the information bits.

The mapper $\Phi$ is defined via the input alphabet $\mX=[\bx_0^\T,\ld,\bx_{M-1}^\T]^\T\in \R^{M\times N}$, where $m$ bits are used to index the $M= 2^m$ symbols vectors $\bx_i \in \R^N$ for $i=0,\ldots,M-1$. We associate with each symbol $\bx_i$ the codeword (binary labeling) $\bc_i \triangleq [c_{i,0},\ldots,c_{i,m-1}] \in\mcB^m$ and the probability $0\le P_i\le 1$, where $P_i\triangleq P_{\bX}(\bx_i)$. The binary labeling is denoted by $\mL = [\bc_0^\T,\ldots,\bc_{M-1}^\T]^\T \in \mcB^{M\times m}$ and the input distribution by $\mP = [P_0,\ldots,P_{M-1}]^\T \in [0,1]^M$.

In the following, the labeling used throughout this paper is defined. This can be done without loss of generality, as will be explained in Sec.~\ref{Sec:preliminaries:Shaping}.
\begin{definition}[Natural binary code] \label{def-nbc}
The natural binary code (NBC) is the binary labeling $\mN_m \triangleq [\bn(0)^\T,\ldots,\bn(M-1)^\T]^\T$, where $\bn(i) = [n_{i,0},\ldots,n_{i,m-1}] \in \mcB^m$ denotes the base-2 representation of the integer $0\le i \le M-1$, with $n_{i,m-1}$ being the most significant bit.
\end{definition}

This definition of the NBC is different from the one in \cite{Agrell10b}. The difference lies only in the bit ordering, i.e., in this paper we consider the last column of $\mN_m$ to contain the most significant bits of the base-2 representation of the integers $i=0,1,\ld,M-1$.
It follows from Definition~\ref{def-nbc} that 
\eqlab{nbc-poweroftwo}{
n_{2^l,k} = \begin{cases}
1, & k=l, \\
0, & k\ne l
\end{cases}
}
for $k=0,\ldots,m-1$ and $l=0,\ldots,m-1$, and
\eqlab{nbc-oplus}{
n_{i\oplus j,k} = n_{i,k}\oplus n_{j,k}
}
for $i=0,\ldots,M-1$, $j=0,\ldots,M-1$ and $k=0,\ldots,m-1$.

\subsection{Probabilistic Shaping in BICM}\label{Sec:preliminaries:Shaping}

Assuming independent, but possibly nonuniformly distributed, bits $C_0,\ldots,C_{m-1}$ at the input of the modulator (cf.~Fig.~\ref{model}), the symbol probabilities are given by \cite[eq. (30)]{Agrell10b} \cite[eq.~(8)]{Fabregas10a} \cite[eq.~(9)]{Nguyen11}
\eq{
P_i = \prod_{k=0}^{m-1} P_{C_k}(c_{i,k})
}
for $i=0,\ldots,M-1$, where $P_{C_k}(u)$ for $u\in\mcB$ is the probability of $C_k=u$. Since $P_{C_k}(1)=1-P_{C_k}(0)$, the distribution $\mP$ is fully specified by the vector of bit probabilities $\bb\triangleq [P_{C_0}(0),\ld,P_{C_{m-1}}(0)]$.

Throughout this paper, we assume that $0<P_{C_k}(0) <1$ for all $k=0,\ldots,m-1$; i.e., all constellation points are used with a nonzero probability. This can be done without loss of generality, because if $P_{C_k}(0) = 0$ or $P_{C_k}(0) = 1$ for some $k$, then half of the constellation points will never be transmitted. If this is the case, the corresponding branches in Fig.~\ref{model} are removed, $m$ is reduced by one, and the mapper $\Phi$ is redefined accordingly.\footnote{Constellations with $P_{C_k}=0$ for some $k$ can yield counter-intuitive results, such as Gray-labeled constellations being FOO (see \cite{Fabregas10a,Peng12} and Example~\ref{8PAM.Example}.)}
The result is another BICM scheme with identical performance, which satisfies $0<P_{C_k}(0) <1$ for all $k$. 

For any constellation $[\mX, \mP, \mL]$, a set of equivalent constellations can be constructed by permuting the rows of $\mX$, $\mL$, and $\mP$, provided that the same permutation is applied to all three matrices. Specifically, denote the permutation that maps the NBC into the desired labeling $\mL$ by $\Pi$, i.e., $\Pi(\mN_m)=\mL$. The BICM system defined by the alphabet $\Pi(\mX)$, the distribution $\Pi(\mP)$, and the labeling $\Pi(\mN_m)=\mL$ is entirely equivalent to the system with alphabet $\mX$, distribution $\mP$, and labeling $\mN_m$. Without loss of generality, the analysis in this paper is therefore restricted to the latter case.

Based on the previous discussion, from now on we use the name \emph{constellation} to denote the pair $[\mX, \mP]$, where the NBC labeling is implicit. Thus, $\mL = \mN_m$ and $c_{i,k}=n_{i,k}$ for all $i$ and $k$, which simplifies the analysis. Note that $\mP$ cannot be chosen arbitrarily in BICM; only distributions that satisfy
\eqlab{pck-prod}{
P_i = \prod_{k=0}^{m-1} P_{C_k}(n_{i,k})
}
for some vector of bit probabilities $\bb$ will be considered in the paper. An important special case is the \emph{uniform distribution}, for which $\bb = [1/2,\ldots,1/2]$ and $\mP = \mU_m \triangleq [1/M, \ldots,1/M]^\T$.

\subsection{The Hadamard Transform}\label{Sec:preliminaries:HT}

The Hadamard transform (HT), or Walsh--Hadamard transform, is a discrete, linear, orthogonal transform, whose coefficients take values in $\pm 1$. It is popular in image processing \cite{Pratt69} and can be used to analyze various aspects of binary labelings in digital communications and source coding \cite{Calderbank84,Knagenhjelm96,Khan10,Agrell10b}.

\begin{definition}
The HT $\tX = [\tbx_0^\T,\ldots,\tbx_{M-1}^\T]^\T$ of a matrix (or vector) $\mX = [\bx_0^\T,\ldots,\bx_{M-1}^\T]^\T$ with $M=2^m$ rows is
\eqlab{HTdef}{
\tbx_i \triangleq \frac{1}{M}\sum_{j=0}^{M-1} \bx_j h_{i,j}, \qquad i=0,\ldots,M-1
}
where for all $i=0,\ldots,M-1$ and $j=0,\ldots,M-1$
\eqlab{hcoeff}{
h_{i,j} = \prod_{k=0}^{m-1} (-1)^{n_{i,k} n_{j,k}}
.}
\end{definition}

Because $n_{0,k} = 0$ for $k=0,\ldots,m-1$, setting $i=0$ in \eqref{HTdef}--\eqref{hcoeff} shows that the first HT vector
\eqlab{HT0}{
\tbx_0 = \frac{1}{M}\sum_{j=0}^{M-1} \bx_j
}
can be interpreted as the uniformly weighted mean of the alphabet. This is a property that the HT shares with, e.g., the discrete Fourier transform.

It can be shown from \eqref{hcoeff} that
\eqlab{h.ortho}{
\sum_{i=0}^{M-1} h_{i,l} h_{i,j} &= \begin{cases}
M, & j=l, \\ 0, & j \ne l
\end{cases}
}
for all $j=0,\ldots,M-1$ and $l=0,\ldots,M-1$. Therefore, the inverse transform is identical to the forward transform, apart from a scale factor:
\eqlab{HTinv}{
\bx_j = \sum_{i=0}^{M-1} \tbx_i h_{i,j}, \qquad j=0,\ldots,M-1.
}

\subsection{A New Transform}\label{Sec:preliminaries:new-transform}

In this section, we define a linear transform between vectors or matrices, which depends on the input distribution $\mP$ via the bit probabilities $\bb$. Its usage will become clear in Section~\ref{Sec:capacity.capacity-nonuniform}.

\begin{definition} \label{new-transform}
Given the bit probabilities $\bb=[P_{C_0}(0),\ldots,P_{C_{m-1}}(0)]$, the transform $\ntX = [\ntbx_0^\T,\ldots,\ntbx_{M-1}^\T]^\T$ of a matrix (or vector) $\mX = [\bx_0^\T,\ldots,\bx_{M-1}^\T]^\T$ is
\eqlab{transform}{
\ntbx_i \triangleq \sum_{j=0}^{M-1} \bx_j \gamma_{i,j} \sqrt{P_j}, \qquad i=0,\ldots,M-1
}
where $P_j$ is given by \eqref{pck-prod}. The coefficients $\gamma_{i,j}$ are defined as
\eqlab{gammadef}{
\gamma_{i,j} &\triangleq \prod_{k=0}^{m-1}
	\Big[(-1)^{\bar{n}_{i,k}n_{j,k}}\sqrt{P_{C_k}(0)}+(-1)^{n_{i,k}\bar{n}_{j,k}}\sqrt{P_{C_k}(1)}\Big]
}
for all $i=0,\ldots,M-1$ and $j=0,\ldots,M-1$, where the bars represent negation ($\bar{b}=1-b$, see Sec.~\ref{Sec:preliminaries:Notation}).
\end{definition}

\begin{remark} \label{rem.identity}
For equally likely symbols, i.e., $\mP=\mU_m$, the transform becomes the identity operation $\ntX = \mX$, because then $\gamma_{i,i} = \sqrt{M}$ for $i=1,\ldots,M$ and $\gamma_{i,j} = 0$ for $i \ne j$.
\end{remark}

The transform coefficients $\gamma_{i,j}$ are nonsymmetric in the sense that in general $\gamma_{i,j} \ne \gamma_{j,i}$. They have some appealing properties given by the following lemma, which will be used in the proofs of Theorems~\ref{th-inverse}, \ref{thD}, and \ref{thB}.

\begin{lemma}\label{lemmaG}
For any $j=0,\ldots,M-1$ and $l=0,\ldots,M-1$,
\eqlab{G2}{
\sum_{i=0}^{M-1} \gamma_{i,l} \gamma_{i,j} &= \begin{cases}
M, & j=l, \\ 0, & j \ne l,
\end{cases} \\
\sumi h_{l,i}\gamma_{i,j} &= M h_{j,l} \sqrt{P_{j \oplus l}} \label{G4}
}
where $P_j$ is given by \eqref{pck-prod} and $h_{l,i}$ is defined in \eqref{hcoeff}.
\end{lemma}

\begin{IEEEproof}
See the Appendix.
\end{IEEEproof}

We pay particular attention to two important special cases of \eqref{G4}. First, if $l=0$, then 
$h_{l,j} = h_{j,l} = 1$ and $P_{j \oplus l} = P_j$ for $j=0,\ldots,M-1$. Second, if $l=2^k$ for any integer $k=0,\ldots,m-1$, then by \eqref{hcoeff}, $h_{l,i} = h_{i,l} = (-1)^{n_{i,k}}$ for any $i=0,\ldots,M-1$ and by \eqref{pck-prod}
\eq{
P_{j \oplus l} = \prod_{k'=0}^{m-1}\pck(n_{j\oplus 2^k,k'})
.}
Using first \eqref{nbc-oplus} and then \eqref{nbc-poweroftwo}, we obtain
\eq{
P_{j \oplus l} &= \left(\prod_{k'=0}^{m-1} \pck(n_{j,k'})\right) \frac{\pck(n_{j,k} \oplus 1)}{\pck(n_{j,k})} \\
&= P_j\frac{\pck(\bar{n}_{j,k})}{\pck(n_{j,k})}
.}
Substituting these two cases ($l=0$ and $l=2^k$) into \eqref{G4} proves the following corollary.
\begin{corollary}\label{corG}
For any $j=0,\ldots,M-1$,
\eqlab{G1}{
\sum_{i=0}^{M-1} \gamma_{i,j} & = M\sqrt{P_j} \\
\sumi (-1)^{n_{i,k}} \gamma_{i,j} &= M (-1)^{n_{j,k}} \sqrt{P_j\frac{\pck(\bar{n}_{j,k})}{\pck(n_{j,k})}} \label{G3}
.}
\end{corollary}

The fact that the sums $\sumi \gamma_{i,l}\gamma_{i,j}$ in \eqref{G2} are zero whenever $j\ne l$, independently of the input distribution, implies that the coefficients $\gamma_{i,j}$ form an orthogonal basis. As a consequence, the transform is invertible, as shown in the next theorem.

\begin{theorem}\label{th-inverse}
The inverse transform $\mX = [\bx_0^\T,\ldots,\bx_{M-1}^\T]^\T$ of a matrix (or vector) $\ntX = [\ntbx_0^\T,\ldots,\ntbx_{M-1}^\T]^\T$ is, given the bit probabilities $\bb=[P_{C_0}(0),\ldots,P_{C_{m-1}}(0)]$,
\eqlab{inverse-transform}{
\bx_j = \frac{1}{M\sqrt{P_j}} \sum_{i=0}^{M-1} \ntbx_i \gamma_{i,j}, \qquad j=0,\ldots,M-1.
}
\end{theorem}

\begin{IEEEproof}
For $j=0,\ldots,M-1$,
\eq{
\sum_{i=0}^{M-1} \ntbx_i \gamma_{i,j}
	&= \sum_{i=0}^{M-1} \gamma_{i,j} \sum_{l=0}^{M-1} \bx_l \gamma_{i,l} \sqrt{P_l} \\
	&= \sum_{l=0}^{M-1} \bx_l \sqrt{P_l} \sum_{i=0}^{M-1} \gamma_{i,l} \gamma_{i,j}
.}
Applying \eqref{G2} and dividing both sides by $M \sqrt{P_j}$, which by Sec.~\ref{Sec:preliminaries:Shaping} is nonzero, completes the proof.
\end{IEEEproof}

\begin{example}\label{G.Example}
If the bit probabilities are $\bb=[0.35,0.50]$, then the symbol probabilities \eqref{pck-prod} are $\mP = [0.175,0.325,0.175,0.325]^\T$. The transform coefficients $\gamma_{i,j}$ in \eqref{gammadef} are the elements at row $i$, column $j$ of
\eqlab{Gmatrix}{
\mG = \begin{bmatrix}
1.977 & 0.304 & 0 & 0 \\
-0.304 & 1.977 & 0 & 0 \\
0 & 0 & 1.977 & 0.304 \\
0 & 0 & -0.304 & 1.977
\end{bmatrix}
}
It is readily verified that $\mG^\T \mG = M \mI$, which is \eqref{G2} in matrix notation. The mean values in each column of \eqref{Gmatrix} are $[0.418,0.570,0.418,0.570]^\T$, which in agreement with \eqref{G1} are the square roots of the elements in $\mP$.
Similarly, it can be shown that $\mG$ in \eqref{Gmatrix} satisfies \eqref{G4} and \eqref{G3}.

If the Gray-labeled $4$-ary pulse amplitude modulation (PAM) constellation $[\mX,\mP]$ is considered, $\mX = [-3,-1,3,1]^\T$. Rewriting \eqref{transform} in matrix notation, the transform can be calculated as $\ntX = \mG \mD^{1/2} \mX = [-2.654,-0.746,2.654,0.746]$, where $\mD \triangleq \diag(\mP)$. This nonequally spaced 4-PAM alphabet will be illustrated and analyzed in Example~\ref{Example.16QAM-BRGC}. The inverse transform \eqref{inverse-transform} can be written as $\mX = (1/M)\mD^{-1/2} \mG^\T \ntX$. For a uniform distribution, $\mG = \mD^{-1/2}=\sqrt{M}\mI$, which agrees with Remark~\ref{rem.identity}.
\end{example}

\begin{figure}
\begin{center}
\fbox{
\begin{tabular}{c@{\hspace{5mm}}c@{\hspace{5mm}}c}
 & \small Def.~\ref{new-transform}\\[-.5ex]
$\mX$ & $\Longleftrightarrow$ & $\mS=\ntX=\mG \mD^{1/2} \mX$ \\
 & \small Theorem~\ref{th-inverse} \\
\makebox[0em][r]{\raisebox{1.3ex}{\small HT }}\rotatebox{90}{$\Longleftrightarrow$} & & \makebox[0em][r]{\raisebox{1.3ex}{\small HT }}\rotatebox{90}{$\Longleftrightarrow$} \\
 & \small Theorem~\ref{thD} \\[-.5ex]
$\tX = \frac{1}{M}\mH \mX$ & $\Longleftrightarrow$ & $\tS = \frac{1}{M}\mH \mS = \mT \tX$
\end{tabular}
}
	\caption{The relations between the alphabet $\mX$, its transform $\mS$, and their respective Hadamard transforms $\tX$ and $\tS$. The transform matrices $\mG$, $\mD$, $\mH$, and $\mT$ are defined in Examples \ref{G.Example} and \ref{G.Example.3}.}
   \label{visual_transform}
\end{center}
\end{figure}

In Sec.~\ref{Sec:FOOBICM.NUD}, we will need to apply the HT and the new transform after each other to the same alphabet. However, the two transforms do not commute, and the result will therefore depend on in which order the transforms are applied. Of particular interest for our analysis is the setup in Fig.~\ref{visual_transform}, where $\mX$ and $\mS$ are related via the transform defined above. Their HTs $\tX$ and $\tS$ are however not related via the same transform. Instead, a relation between $\tX$ and $\tS$ can be established via the following theorem.

\begin{theorem}\label{thD}
If $\mS=\ntX$, then their HTs $\tilde{\mS}$ and $\tX$ satisfy
\eqlab{eq:thD1}{
\tbs_i &= \psi_i \sum_{j=0}^{M-1}\tbx_j
	\prod_{\substack{k=0\\n_{j,k}\ne n_{i,k}}}^{m-1}\left(P_{C_k}(0)-P_{C_k}(n_{j,k})\right), \nonumber\\
	&\hspace{12em} i=0,\ldots,M-1,\\
\tbx_j &= \sum_{i=0}^{M-1} \frac{\tbs_i}{\psi_i}
	\prod_{\substack{k=0\\n_{i,k}\ne n_{j,k}}}^{m-1}\left(P_{C_k}(n_{i,k})-P_{C_k}(0)\right), \nonumber\\
	&\hspace{12em} j=0,\ldots,M-1 \label{eq:thD2}
}
where
\eqlab{psi}{
\psi_i \triangleq \prod_{\substack{k=0\\n_{i,k}=1}}^{m-1}2\sqrt{P_{C_k}(0)P_{C_k}(1)},
	\qquad i=0,\ldots,M-1
}
and a product over $\varnothing$ is defined as 1.
\end{theorem}

\begin{IEEEproof}
See the Appendix.
\end{IEEEproof}

\begin{remark}\label{triangular}
The summation in \eqref{eq:thD1} can be confined to $\sum_{j=i}^{M-1}$, because whenever $j<i$, there exists at least one bit position $k$ for which $n_{i,k} \ne n_{j,k} = 0$. Analogously, the summation in \eqref{eq:thD2} can be confined to $\sum_{i=j}^{M-1}$.\end{remark}

\begin{example}\label{G.Example.3}
Expression \eqref{eq:thD1} can be written as $\tS = \mT \tX$, or $\tX = \mT^{-1} \tS$. The element at row $i$, column $j$ of $\mT$ and $\mT^{-1}$ are given by \eqref{eq:thD1}--\eqref{eq:thD2} as, resp., $\psi_i \prod_{k:\; n_{j,k}\ne n_{i,k}} (P_{C_k}(0)-P_{C_k}(n_{j,k}))$ and $(1/\psi_j) \prod_{k:\; n_{j,k}\ne n_{i,k}} (P_{C_k}(n_{j,k})-P_{C_k}(0))$. With $\bb$, $\mX$ and $\ntX$ from Example~\ref{G.Example}, we obtain $[\psi_0,\ldots,\psi_3] = [1,0.954,1,0.954]$ and
\eqlab{Tmatrix}{
\mT &= \begin{bmatrix}
1 & -0.300 & 0 & 0 \\
0 & 0.954 & 0 & 0 \\
0 & 0 & 1 & -0.300 \\
0 & 0 & 0 & 0.954
\end{bmatrix}, \;
\mT^{-1} = \begin{bmatrix}
1 & 0.315 & 0 & 0 \\
0 & 1.048 & 0 & 0 \\
0 & 0 & 1 & 0.315 \\
0 & 0 & 0 & 1.048
\end{bmatrix}
}
which, as predicted by Remark~\ref{triangular}, are upper triangular.

Another relation between $\tX$ and $\tS$ can be deduced from Fig.~\ref{visual_transform}. Defining the Hadamard matrix $\mH$ as the matrix with elements $h_{i,j}$ for $i,j = 0,\ldots,M-1$, the HT relations \eqref{HTdef} and \eqref{HTinv} yield $\tS = (1/M) \mH \mS$ and $\mX = \mH \tX$. Since from Example~\ref{G.Example} $\mS = \ntX = \mG \mD^{1/2} \mX$, we conclude that $\tS = (1/M) \mH \mG \mD^{1/2} \mH \tX$, which implies that $\mT = (1/M) \mH \mG \mD^{1/2} \mH$. Because $\mH^{-1} = (1/M)\mH$ (see \eqref{h.ortho}) and $\mG^{-1} = (1/M)\mG^\T$, the inverse relation is $\mT ^{-1}= (1/M^2) \mH \mD^{-1/2} \mG^\T \mH$. It is straightforward to verify that $\mT$ and $\mT^{-1}$ calculated in this manner, using the numerical values of $\mG$ and $\mD$ in Example~\ref{G.Example}, indeed yield \eqref{Tmatrix}.
\end{example}

\section{BICM at low SNR}\label{Sec:capacity}

\subsection{Mutual Information}\label{Sec:capacity.MI}

The \emph{mutual information} (MI) in bits per channel use between the random vectors $\bX$ and $\bY$ for an arbitrary channel parameter $\bH$ perfectly known at the receiver is defined as
\eq{
I(\bX;\bY|\bH)  \triangleq \E\left[\log_2{\frac{p_{\bY|\bX,\bH}(\bY|\bX,\bH)}{p_{\bY|\bH}(\bY|\bH)}}\right]
}
where the expectation is taken over the joint pdf $p_{\bX,\bY,\bH}$, and $p_{\bY|\bX,\bH}$ is given by \eqref{ctpdf_AWGN}.

The MI between $\bX$ and $\bY$ conditioned on the value of the $k$th bit at the input of the modulator is defined as
\eq{
I(\bX;\bY|\bH,C_k) 	\triangleq \E\left[\log_2{\frac{p_{\bY|\bX,\bH,C_k}(\bY|\bX,\bH,C_k)}{p_{\bY|\bH,C_k}(\bY|\bH,C_k)}}\right]
}
where the expectation is taken over the joint pdf $p_{\bX,\bY,\bH,C_k}$.

\begin{definition}[BICM Generalized Mutual Information]
The BICM generalized mutual information (BICM-GMI) is defined as \cite{Caire98,Martinez08b,Martinez09,Stierstorfer09a}
\begin{align}\label{GMI.BI}
\MI\left(\SNR\right) &\triangleq \sum_{k=0}^{m-1} I(C_k;\bY|\bH) \nonumber\\
&= mI(\bX;\bY|\bH)-
\sum_{k=0}^{m-1}I(\bX;\bY|\bH,C_k)
\end{align}
\end{definition}
where the second line follows by the chain rule. We will analyze the right-hand side of \eqref{GMI.BI} as a function of $\SNR$, for a given pdf $p_H$. According to \eqref{snr.definition}, $\SNR$ can be varied in two ways, either by varying $N_0$ for a fixed constellation $[\mX,\mP]$ or, equivalently, by rescaling the alphabet $\mX$ linearly for fixed $N_0$ and input distribution $\mP$.

Martinez \emph{et al.} \cite{Martinez09} recognized the BICM decoder in Fig.~\ref{model} as a mismatched decoder and showed that the BICM-GMI in \eqref{GMI.BI} corresponds to an achievable rate of such a decoder. This means that reliable transmission using a BICM system at rate $\Rc$ is possible if $\Rc\leq \MI\left(\SNR\right)$. Since from \eqref{snr.definition} $\Ebr/N_0 = \SNR/\Rc$, the inequality $\Rc\leq I(\SNR)$ gives\footnote{The definition of the related function $f(\Rc)$ in \cite[eq.~(37)]{Agrell10b} is erroneous and should read ``$E_\text{b}/N_0$ is bounded from below by $f(\Rc)/\E_{H}[H^2]$, where $f(\Rc) \triangleq \mf{C}^{-1}(\Rc)/\Rc$.''}
\eqlab{min.EbN0-anySNR}{
\frac{\Ebr}{N_0} \ge \frac{\SNR}{I(\SNR)}
}
for any $\SNR$. Focusing on the wideband regime, i.e., asymptotically low SNR, we make the following definition.

\begin{definition}[Low-GMI Parameters] \label{def-cap-par}
The \emph{low-GMI parameters} of a constellation $[\mX, \mP]$ are defined as $\left[\bmu,\Es,\alpha\right]$, where
\eq{
\bmu &\triangleq \E[\bX] \\
\Es &\triangleq \E[\|\bX\|^2] \\
\alpha &\triangleq \left.\frac{d \MI(\SNR)}{d \SNR} \right|_{\SNR = 0}
.}
\end{definition}

In the wideband regime, the average bit energy-to-noise ratio needed for reliable transmission is, using \eqref{min.EbN0-anySNR} and the definition of $\alpha$, lower-bounded by
\begin{align}\label{min.EbN0}
\frac{\Ebr}{N_0} \geq \lim_{\SNR\rightarrow 0^+} \frac{\SNR}{\MI(\SNR)}
= \frac{1}{\alpha}
.\end{align}
Furthermore, since in the wideband regime ${\Ebr}/{N_0} \geq \log_\text{e}2 = -1.59$ dB \cite{Verdu02}, $\alpha^{-1} \ge -1.59$ dB.

The first-order behavior of the BICM-GMI in \eqref{GMI.BI} is fully determined by $\alpha$, which, as we shall see later (e.g., in \eqref{thAproof}), in turn depends on $\bmu$ and $\Es$. This is why we designate this triplet as low-GMI parameters. The same definitions can be applied to other MI functions $\MI\left(\SNR\right)$ such as the coded modulation MI (CM-MI) \cite{Agrell10b}. In this paper, however, we are only interested in the BICM-GMI.

The main contributions of this paper are to characterize the low-GMI parameters for arbitrary constellations, including those with nonuniform distributions (Sec.~\ref{Sec:capacity.capacity-nonuniform}), and to identify the set of constellations for BICM that maximize $\alpha$, i.e., minimize $\Ebr/{N_0}$ in the wideband regime (Sec.~\ref{Sec:FOOBICM.NUD}).

\subsection{Low-GMI Parameters for Uniform Distributions}\label{Sec:capacity.Uniform}

The low-GMI parameters $[\bmu, \Es, \alpha]$ have been analyzed in detail for arbitrary input alphabets $\mX$ under the assumption of uniform probabilities \cite{Agrell10b}. Under this assumption, they can be expressed as given by the following theorem.

\begin{theorem}\label{thAuni}
For a constellation $[\mX,\mU_m]$, the low-GMI parameters are
\eqlab{A01}{
\bmu &= \frac{1}{M}\sum_{i=0}^{M-1} \bx_i, \\
\Es &= \frac{1}{M} \sum_{i=0}^{M-1} \|\bx_i\|^2, \label{A02}\\
\alpha &= \frac{\log_2\text{e}}{M^2 \Es} \sum_{k=0}^{m-1}
	\left\| \sum_{i=0}^{M-1} (-1)^{n_{i,k}} \bx_i \right\|^2. \label{A03}
}
\end{theorem}

\begin{IEEEproof}
Expressions \eqref{A01} and \eqref{A02} follow directly from Definition~\ref{def-cap-par}, while \eqref{A03} was proved in \cite[eq.~(50)]{Agrell10b}.
\end{IEEEproof}

The low-GMI parameters can be conveniently expressed as functions of the HT $\tX$ of the alphabet $\mX$, as shown in the following theorem.

\begin{theorem}\label{thC}
The low-GMI parameters can be expressed as
\begin{align}
\bmu &= \tbx_0, \label{thC.1}\\
\Es &= \sum_{i=0}^{M-1} \|\tbx_i\|^2, \label{thC.2}\\
\alpha &= \frac{\log_2\text{e}}{\Es}\sum_{k=0}^{m-1} \|\tbx_{2^k}\|^2. \label{thC.3}
\end{align}
\end{theorem}
\begin{IEEEproof}
The expression \eqref{thC.1} is obtained from \eqref{HT0}, \eqref{thC.2} from \cite[eq.~(16)]{Agrell10b}, and \eqref{thC.3} from \cite[Th.~11]{Agrell10b}.
\end{IEEEproof}

\subsection{Low-GMI Parameters for Nonuniform Distributions}\label{Sec:capacity.capacity-nonuniform}

\begin{table*}
\caption{Low-GMI parameters and FOO conditions for BICM using uniform and nonuniform input distributions. The results for $[\mX,\mU_m]$ are from \cite{Agrell10b} (cf.~Theorems~\ref{thC} and \ref{thC1}) and the ones for $[\mX,\mP]$ or $[\ntX,\mU_m]$ are from Theorems~\ref{thA}, \ref{thB}, and \ref{thF}.}
\renewcommand{\arraystretch}{1.5}
\begin{center}
\begin{tabular}{ccc}
\hline
				& $[\mX,\mU_m]$ 							& $[\mX,\mP]$ or $[\ntX,\mU_m]$ 	\\
\hline

\hline
	$\bmu$ 		& $\tbx_0$										& $\displaystyle{\sum_{i=0}^{M-1}} P_i \bx_i$ \\
	$\Es$  		& $\displaystyle{\sum_{i=0}^{M-1} \|\tbx_i\|^2}$ 						& $\displaystyle{\sum_{i=0}^{M-1}} P_i \|\bx_i\|^2$\\
	$\alpha$		& $\dfrac{\log_2\text{e}}{\Es}\displaystyle{\sum_{k=0}^{m-1}} \|\tbx_{2^k}\|^2$	& $\frac{\log_2{\text{e}}}{\Es} \displaystyle{\sum_{i=0}^{M-1}} P_i \sum_{j=0}^{M-1} P_j \inner{\bx_i}{\bx_j}\sum_{k=0}^{m-1}(-1)^{n_{i,k}+n_{j,k}} \dfrac{P_{C_k}(\bar{n}_{i,k})}{P_{C_k}(n_{j,k})}$	\\
\hline
	FOO Condition 	& $\tbx_j = \bzero$, $\forall j\notin\set{1,2,4,\ldots,M/2}$					& $\bmu = \bzero$ and $\tbx_j = \bzero$, $\forall j\notin\set{0} \cup \set{1,2,4,\ldots,M/2}$
\\	
\hline
\end{tabular}
\end{center}
\label{summary}
\end{table*}

The next theorem is analogous to Theorem \ref{thAuni} but applies to an arbitrary input distribution.

\begin{theorem}\label{thA}
For a constellation $[\mX,\mP]$, the low-GMI parameters are
\eqlab{thA1}{
\bmu &= \sum_{i=0}^{M-1} P_i \bx_i, \\
\Es &= \sum_{i=0}^{M-1} P_i \|\bx_i\|^2, \label{thA2}\\
\alpha &= \frac{\log_2{\text{e}}}{\Es} \sum_{i=0}^{M-1} P_i \sum_{j=0}^{M-1} P_j \inner{\bx_i}{\bx_j} \nonumber\\
&\qquad\qquad\qquad\cdot  \sum_{k=0}^{m-1}
(-1)^{n_{i,k}+n_{j,k}} \frac{P_{C_k}(\bar{n}_{i,k})}{P_{C_k}(n_{j,k})}. \label{thA3}
}
\end{theorem}

\begin{IEEEproof}
Again, \eqref{thA1} and \eqref{thA2} follow from Definition \ref{def-cap-par}, while \eqref{thA3} requires some analysis. It was shown in \cite[Th.~10]{Agrell10b} that
\eqlab{thAproof}{
\alpha &= \frac{\log_2{\text{e}}}{2\Es} \sum_{k=0}^{m-1} \Bigg[\left\|\sum_{i=0}^{M-1} (-1)^{n_{i,k}}\frac{P_i \bx_i}{\sqrt{P_{C_k}(n_{i,k})}} \right\|^2 \nonumber\\
&\qquad+\left\|\sum_{i=0}^{M-1} \frac{P_i \bx_i}{\sqrt{P_{C_k}(n_{i,k})}} \right\|^2-2 \|\bmu\|^2\Bigg]
.}
Substituting \eqref{thA1} and writing the squared norms as the inner products of two identical vectors yields
\eq{
\alpha &= \frac{\log_2{\text{e}}}{2\Es} \sum_{k=0}^{m-1} \sum_{i=0}^{M-1} \sum_{j=0}^{M-1}
P_i P_j \inner{\bx_i }{\bx_j} \nonumber\\
&\qquad\cdot \Bigg[
\frac{(-1)^{n_{i,k}}(-1)^{n_{j,k}}}{\sqrt{P_{C_k}(n_{i,k})P_{C_k}(n_{j,k})}} \nonumber\\
&\qquad\qquad+\frac{1}{\sqrt{P_{C_k}(n_{i,k})P_{C_k}(n_{j,k})}}
-2
\Bigg]
.}
The expression in brackets can be simplified as
\eq{
\frac{(-1)^{n_{i,k}+n_{j,k}}+1}{\sqrt{P_{C_k}(n_{i,k})P_{C_k}(n_{j,k})}} -2 &= \begin{cases}
2\frac{P_{C_k}(\bar{n}_{j,k})}{P_{C_k}(n_{j,k})},  &n_{j,k} = n_{i,k} \\
-2, &n_{j,k} \ne n_{i,k}
\end{cases} \\
&= 2 (-1)^{n_{i,k}+n_{j,k}}\frac{P_{C_k}(\bar{n}_{i,k})}{P_{C_k}(n_{j,k})}
}
which completes the proof of \eqref{thA3}.
\end{IEEEproof}

Theorem~\ref{thA} shows that the low-GMI parameters depend on the input alphabet $\mX$, the binary labeling (via $n_{i,k}$ in the expression for $\alpha$), and the input distribution (via $P_{C_k}(u)$ and $P_i$). While the low-GMI parameters of an alphabet $\mX$ with uniform probabilities are conveniently expressed in terms of its HT $\tX$ (cf.~Theorem~\ref{thC}), no similar expressions are known for the low-GMI parameters of a general constellation in \eqref{thA1}--\eqref{thA3}. This has so far prevented the analytic optimization of such constellations. The new transform introduced in Section~\ref{Sec:preliminaries:new-transform}, however, solves this problem by establishing an equivalence between an arbitrary constellation, possibly with nonuniform probabilities, and another constellation with uniform probabilities.

\begin{theorem}\label{thB}
The low-GMI parameters $[\bmu,\Es,\alpha]$ of any constellation $[\mX,\mP]$ are equal to the low-GMI parameters of $[\ntX,\mU_m]$.
\end{theorem}

\begin{IEEEproof}
Let the low-GMI parameters of $[\ntX,\mU_m]$ be denoted by $[\bmu',\Es',\alpha']$. First, \eqref{A01} and \eqref{transform} yield
\eqlab{bmuprime}{
\bmu' 	&= \frac{1}{M} \sum_{i=0}^{M-1} \sum_{j=0}^{M-1} \bx_j \gamma_{i,j} \sqrt{P_j} \nonumber\\
				&= \frac{1}{M} \sum_{j=0}^{M-1} \bx_j \sqrt{P_j} \sum_{i=0}^{M-1} \gamma_{i,j}
.}
Applying \eqref{G1} to the inner sum in \eqref{bmuprime} reveals that
\eq{
\bmu' 	&= \sum_{j=0}^{M-1} P_j \bx_j = \bmu
.}

Second, \eqref{A02} and \eqref{transform} yield
\eq{
\Es' &= \frac{1}{M} \sum_{i=0}^{M-1} \| \bx_i \|^2 \\
	&= \frac{1}{M} \sum_{i=0}^{M-1} \inner{\sum_{j=0}^{M-1} \bx_j \gamma_{i,j} \sqrt{P_j}}
		{\sum_{l=0}^{M-1} \bx_l \gamma_{i,l} \sqrt{P_l}} \\
	&= \frac{1}{M} \sum_{j=0}^{M-1} \sum_{l=0}^{M-1}
		\inner{\bx_j}{\bx_l} \sqrt{P_j P_l}
		\sum_{i=0}^{M-1} \gamma_{i,j} \gamma_{i,l}
.}
Evaluating the inner sum using \eqref{G2} gives
\eq{
\Es' &= \sum_{j=0}^{M-1} P_j \|\bx_j\|^2 = \Es
.}

For the third and last part of the theorem, \eqref{A03} yields
\eqlab{alphap}{
\alpha' = \frac{\log_2\text{e}}{M^2\Es} \sumk \left\| \sumi (-1)^{n_{i,k}} \ntbx_i \right\|^2
}
where $\ntbx_i$ is given by \eqref{transform}. The inner sum can be expanded as
\eq{
\sumi (-1)^{n_{i,k}} \ntbx_i
&= \sumi (-1)^{n_{i,k}} \sumj \bx_j \gamma_{i,j} \sqrt{P_j} \nonumber\\
&= \sumj \bx_j \sqrt{P_j} \sumi (-1)^{n_{i,k}} \gamma_{i,j}
.}
Applying \eqref{G3} to the inner sum, we obtain
\eq{
\sumi (-1)^{n_{i,k}} \ntbx_i = M\sumj \bx_j P_j (-1)^{n_{j,k}} \sqrt{\frac{\pck(\bar{n}_{j,k})}{\pck(n_{j,k})}}
.}
We take the inner product of this vector with itself and substitute the obtained expression for the squared norm in \eqref{alphap}. This yields, after rearranging terms,
\eqlab{alphap2}{
\alpha' &= \frac{\log_2\text{e}}{\Es} \sumk \sumi \sumj \inner{\bx_i}{\bx_j} \nonumber\\
&\qquad\cdot P_i P_j (-1)^{n_{i,k}+n_{j,k}} \sqrt{\frac{\pck(\bar{n}_{i,k})}{\pck(n_{i,k})} \frac{\pck(\bar{n}_{j,k})}{\pck(n_{j,k})}} \\
&= \alpha \label{alphap3}
.}
The square root in \eqref{alphap2} is $\pck(\bar{n}_{i,k})/\pck(n_{i,k})$ if $n_{j,k} = n_{i,k}$ or $1$ if $n_{j,k} = \bar{n}_{i,k}$. In both cases, it can be expressed as $\pck(\bar{n}_{i,k})/\pck(n_{j,k})$ (or, equivalently, $\pck(\bar{n}_{j,k})/\pck(n_{i,k})$). Comparing this result with \eqref{thA3} gives \eqref{alphap3}.
\end{IEEEproof}

Theorem~\ref{thB} shows that the constellation $[\mX,\mP]$ can be mapped to another constellation $[\ntX,\mU_m]$ with the same low-GMI parameters, where $\ntX$ is related to $\mX$ via \eqref{transform} and \eqref{inverse-transform}. This relation between $[\mX,\mP]$ and $[\ntX,\mU_m]$ will be applied in Sec.~\ref{Sec:FOOBICM.NUD} to prove Theorem \ref{thE}, which is the main result of the paper. To summarize, Table~\ref{summary} lists the low-GMI parameters for BICM given by Theorems~\ref{thC} and \ref{thA}. The equivalence of the parameters for $[\mX,\mP]$ or $[\ntX,\mU_m]$ comes from Theorem~\ref{thB}.

\subsection{Numerical Examples}\label{Sec:capacity.capacity-examples}

\begin{figure}
\newcommand{\scale}{0.8}
\setlength{\xs}{1.6mm}
\setlength{\ys}{-2mm}
\psfrag{X0}[cl][cl][\scale]{\overlay{\xs}{\ys}{0}}
\psfrag{X1}[cl][cl][\scale]{\overlay{\xs}{\ys}{1}}
\psfrag{X2}[cl][cl][\scale]{\overlay{\xs}{-\ys}{2}}
\psfrag{X3}[cl][cl][\scale]{\overlay{\xs}{-\ys}{3}}
\psfrag{X4}[cl][cl][\scale]{\overlay{\xs}{\ys}{4}}
\psfrag{X5}[cl][cl][\scale]{\overlay{\xs}{\ys}{5}}
\psfrag{X6}[cl][cl][\scale]{\overlay{\xs}{-\ys}{6}}
\psfrag{X7}[cl][cl][\scale]{\overlay{\xs}{-\ys}{7}}
\psfrag{X8}[cl][cl][\scale]{\overlay{\xs}{\ys}{8}}
\psfrag{X9}[cl][cl][\scale]{\overlay{\xs}{\ys}{9}}
\psfrag{X10}[cl][cl][\scale]{\overlay{\xs}{-\ys}{10}}
\psfrag{X11}[cl][cl][\scale]{\overlay{\xs}{-\ys}{11}}
\psfrag{X12}[cl][cl][\scale]{\overlay{\xs}{\ys}{12}}
\psfrag{X13}[cl][cl][\scale]{\overlay{\xs}{\ys}{13}}
\psfrag{X14}[cl][cl][\scale]{\overlay{\xs}{-\ys}{14}}
\psfrag{X15}[cl][cl][\scale]{\overlay{\xs}{-\ys}{15}}
\psfrag{C0}[cl][cl][\scale]{}
\psfrag{C1}[cl][cl][\scale]{}
\psfrag{C2}[cl][cl][\scale]{}
\psfrag{C3}[cl][cl][\scale]{}
\psfrag{C4}[cl][cl][\scale]{}
\psfrag{C5}[cl][cl][\scale]{}
\psfrag{C6}[cl][cl][\scale]{}
\psfrag{C7}[cl][cl][\scale]{}
\psfrag{C8}[cl][cl][\scale]{}
\psfrag{C9}[cl][cl][\scale]{}
\psfrag{C10}[cl][cl][\scale]{}
\psfrag{C11}[cl][cl][\scale]{}
\psfrag{C12}[cl][cl][\scale]{}
\psfrag{C13}[cl][cl][\scale]{}
\psfrag{C14}[cl][cl][\scale]{}
\psfrag{C15}[cl][cl][\scale]{}
\begin{center}
	\includegraphics{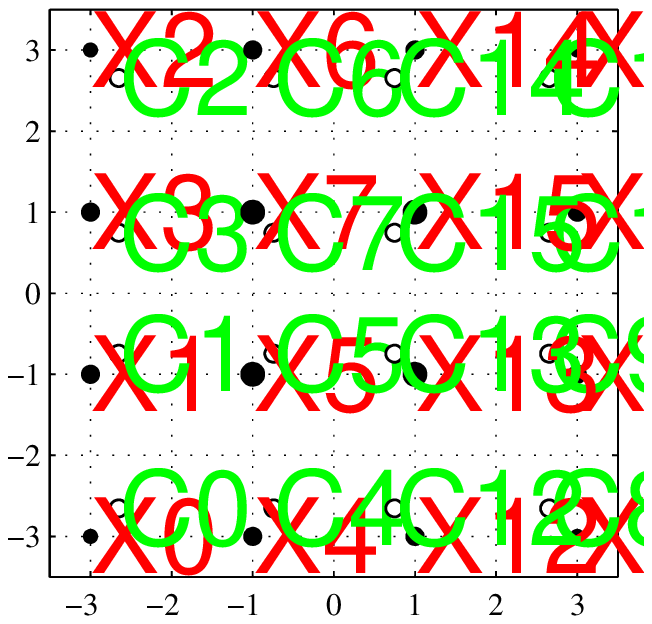}
    \caption{16-QAM constellation $[\mX_\text{QAM},\mP_1]$ (black circles) with bit probabilities $\bb_1=[0.35,0.50,0.35,0.50]$. Each symbol $\bx_j$ is marked with its index $j$, and its probability $P_j$ is proportional to the area of the corresponding circle. White circles represent the transformed constellation $[\ntX_\text{QAM},\mU_4]$, which has the same low-GMI parameters.}
    \label{16QAM_0.350_0.500_0.350_0.500}
\end{center}
\end{figure}

In this Section, we show examples of how the transform defined in Sec.~\ref{Sec:preliminaries:new-transform} works and we also present equivalent constellations $[\mX,\mP]$ and $[\ntX,\mU_m]$. All results are for the AWGN channel. The (G)MIs are numerically evaluated using Gauss--Hermite quadratures following \cite[Sec.~III]{Alvarado11b}.

\begin{example}\label{Example.16QAM-BRGC}
Consider the equally spaced 16-ary square quadrature amplitude modulation (16-QAM) alphabet $\mX=\mX_\text{QAM}$ labeled by the binary reflected Gray code (BRGC) \cite{Agrell04} with bit probabilities $\bb_1=[0.35,0.5,0.35,0.5]$, shown with black circles in Fig.~\ref{16QAM_0.350_0.500_0.350_0.500}. The input distribution $\mP=\mP_1$ given by \eqref{pck-prod} is $P_{0}=P_{2}=P_{8}=P_{10} = 0.031$, $P_{1}=P_{3}= P_{4}=P_{6}=P_{9}=P_{11}=P_{12}=P_{14} = 0.057$, and $P_{5}=P_{7}=P_{13}=P_{15} = 0.106$. These symbol probabilities are indicated in Fig.~\ref{16QAM_0.350_0.500_0.350_0.500}, where the area of the circle representing $\bx_j$ is proportional to the corresponding probability $P_j$.

Another alphabet $\ntX_\text{QAM}$ is obtained by applying the transform in \eqref{transform} to the constellation $[\mX_\text{QAM},\mP_1]$. The white circles in Fig.~\ref{16QAM_0.350_0.500_0.350_0.500} represent the symbols in $\ntX_\text{QAM}$ using a uniform distribution $\mU_4$. The alphabet $\ntX_\text{QAM}$ is still a rectangular 16-QAM constellation, but a nonuniformly spaced one. Every row in Fig.~\ref{16QAM_0.350_0.500_0.350_0.500} can be regarded as a probabilistically shaped (black) or a geometrically shaped (white) 4-PAM constellation; in fact the same 4-PAM constellations as in Example~\ref{G.Example}.

The low-GMI parameters of the two constellations $[\mX_\text{QAM},\mP_1]$ and $[\ntX_\text{QAM},\mU_4]$, given by Theorems~\ref{thA} and \ref{thAuni}, resp., are identical, as predicted by Theorem~\ref{thB}. These are
\begin{align}
\bmu = \bzero,\qquad
\Es = 7.60,\qquad
\label{Example.16QAM-BRGC.alpha}
\alpha = 1.10.
\end{align}

The BICM-GMI for the constellations $[\mX_\text{QAM},\mP_1]$ and $[\ntX_\text{QAM},\mU_4]$ are shown in Fig.~\ref{MI_transformed_QAM_constellations}. In this figure, we also show the capacity of the AWGN channel $\C^\text{AW}$ \cite[eq.~(22)]{Agrell10b} and the CM-MI and BICM-GMI for 16-QAM using a uniform input distribution, i.e., $[\mX_\text{QAM},\mU_4]$. The results show that the BICM-GMIs of the original constellation $[\mX_\text{QAM},\mP_1]$ and the transformed constellation $[\mathring{\mX}_\text{QAM},\mU_4]$ are in general different; however, they converge in the low-SNR regime. The endpoints of the BICM-GMI curves for $[\mX_\text{QAM},\mP_1]$ and $[\mathring{\mX}_\text{QAM},\mU_4]$ are shown with a white circle, whose value follows from \eqref{Example.16QAM-BRGC.alpha} and \eqref{min.EbN0}. The endpoint for the BICM-GMI curve for the constellation $[\mX_\text{QAM},\mU_4]$ is shown with a white square \cite[eq.~(18)]{Martinez08b}.

\begin{figure}
\newcommand{\scale}{0.9}
\psfrag{xlabel}[cc][cc][\scale]{$\Ebr/N_0$~[dB]}
\psfrag{ylabel}[bc][Bc][\scale]{$\Rc$~[bit/symbol]}
\psfrag{AWGNC}[cl][cl][\scale]{$\C^\text{AW}$}
\psfrag{CM-16QAMUn}[cl][cl][\scale]{CM $[\mX_\text{QAM},\mU_4]$}
\psfrag{BICM-16QAMUn}[cl][cl][\scale]{BICM $[\mX_\text{QAM},\mU_4]$}
\psfrag{BICM-16QAMS}[cl][cl][\scale]{BICM $[\mX_\text{QAM},\mP_1]$}
\psfrag{BICM-16QAMT}[cl][cl][\scale]{BICM $[\mathring{\mX}_\text{QAM},\mU_4]$}
\begin{center}
	\includegraphics[width=\columnwidth]{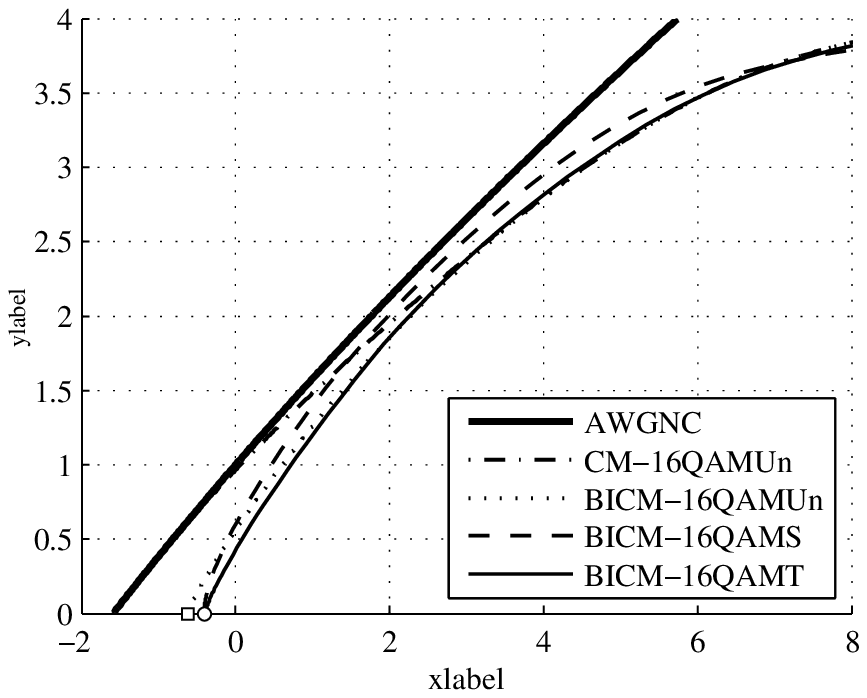}
	\caption{BICM-GMI for probabilistically shaped 16-QAM and its transform. The CM-MI and BICM-GMI for uniform input distributions are also shown, as is the AWGN capacity $\C^\text{AW}$. The white circle and square indicate \eqref{min.EbN0} with $\alpha$ given by \eqref{Example.16QAM-BRGC.alpha} and by \cite[eq.~(55)]{Agrell10b}, resp.}
    \label{MI_transformed_QAM_constellations}
\end{center}
\end{figure}
\end{example}

\begin{figure}
\newcommand{\scale}{.9}
\setlength{\xs}{-0.9mm}
\setlength{\ys}{-1.6mm}
\psfrag{X0}[cl][cl][\scale]{}
\psfrag{X1}[cl][cl][\scale]{}
\psfrag{X2}[cl][cl][\scale]{}
\psfrag{X3}[cl][cl][\scale]{}
\psfrag{X4}[cl][cl][\scale]{}
\psfrag{X5}[cl][cl][\scale]{}
\psfrag{X6}[cl][cl][\scale]{}
\psfrag{X7}[cl][cl][\scale]{}
\psfrag{C0}[cl][cl][\scale]{\overlay{\xs}{-\ys}{0}}
\psfrag{C1}[cl][cl][\scale]{\overlay{\xs}{-\ys}{1}}
\psfrag{C2}[cl][cl][\scale]{\overlay{\xs}{-\ys}{2}}
\psfrag{C3}[cl][cl][\scale]{\overlay{\xs}{-\ys}{3}}
\psfrag{C4}[cl][cl][\scale]{\overlay{\xs}{-\ys}{4}}
\psfrag{C5}[cl][cl][\scale]{\overlay{\xs}{-\ys}{5}}
\psfrag{C6}[cl][cl][\scale]{\overlay{\xs}{-\ys}{6}}
\psfrag{C7}[cl][cl][\scale]{\overlay{\xs}{-\ys}{7}}
\begin{center}
\includegraphics{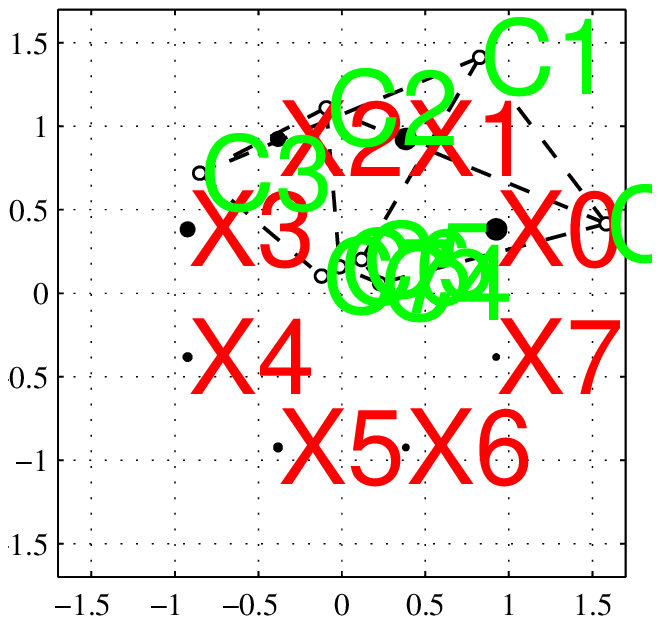}
\end{center}
	\caption{8-PSK constellation $[\mX_\text{PSK},\mP_2]$ with bit probabilities $\bb_2$ (black circles) and the transformed constellation $[\ntX_\text{PSK},\mU_3]$ (white circles), which have the same low-GMI parameters. The circle areas is proportional to the symbol probabilities $P_j$. Dashed lines join symbols whose labels differ in one bit only.}
	\label{8PSK.Constellations.P2}
\end{figure}

\begin{figure}
\newcommand{\scale}{.9}
\setlength{\xs}{-0.9mm}
\setlength{\ys}{-1.6mm}
\psfrag{X0}[cl][cl][\scale]{}
\psfrag{X1}[cl][cl][\scale]{}
\psfrag{X2}[cl][cl][\scale]{}
\psfrag{X3}[cl][cl][\scale]{}
\psfrag{X4}[cl][cl][\scale]{}
\psfrag{X5}[cl][cl][\scale]{}
\psfrag{X6}[cl][cl][\scale]{}
\psfrag{X7}[cl][cl][\scale]{}
\psfrag{C0}[cl][cl][\scale]{\overlay{\xs}{-\ys}{0}}
\psfrag{C1}[cl][cl][\scale]{\overlay{\xs}{-\ys}{1}}
\psfrag{C2}[cl][cl][\scale]{\overlay{\xs}{-\ys}{2}}
\psfrag{C3}[cl][cl][\scale]{\overlay{\xs}{-\ys}{3}}
\psfrag{C4}[cl][cl][\scale]{\overlay{\xs}{-\ys}{4}}
\psfrag{C5}[cl][cl][\scale]{\overlay{\xs}{-\ys}{5}}
\psfrag{C6}[cl][cl][\scale]{\overlay{\xs}{-\ys}{6}}
\psfrag{C7}[cl][cl][\scale]{\overlay{\xs}{-\ys}{7}}
\begin{center}
\includegraphics{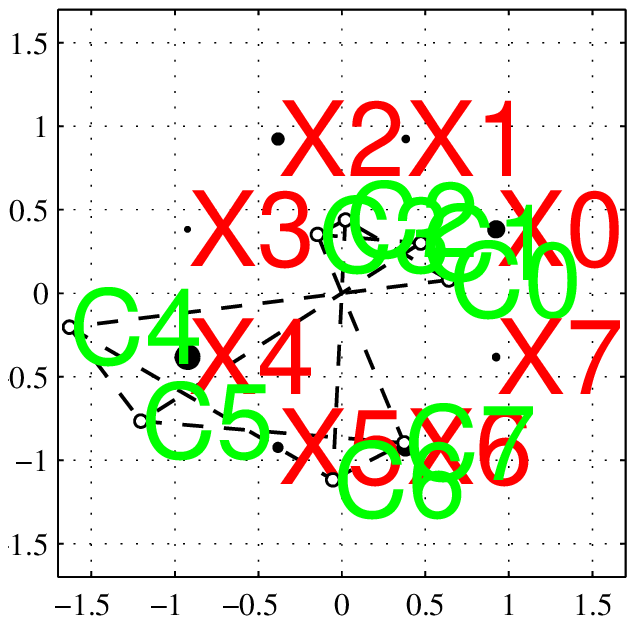}
\end{center}
	\caption{8-PSK constellation $[\mX_\text{PSK},\mP_3]$ with bit probabilities $\bb_3$ (black circles) and its transform $[\ntX_\text{PSK},\mU_3]$ (white circles).}
	\label{8PSK.Constellations.P3}
\end{figure}

\begin{example}\label{Example.8PSK-NBC}
Consider the NBC-labeled $M$-ary phase-shift keying (PSK) alphabet $\mX=\mX_\text{PSK}$, where $\bx_j=[\cos(2\pi j/M+\pi/M),\sin(2\pi j/M+\pi/M)]$ with $j=0,\ld,M-1$. The constellation for $M=8$ for two input distributions $\mP$ are shown in Figs.~\ref{8PSK.Constellations.P2} and \ref{8PSK.Constellations.P3}, where again the circle areas are proportional to the symbol probabilities $P_j$. We denote these input distributions by $\mP_2$ and $\mP_3$, which are generated by $\bb_2=[0.5,0.7,0.9]$ and $\bb_3=[0.9,0.7,0.3]$, resp. The transforms $\ntX_\text{PSK}$ are irregular, not resembling a PSK alphabet. Nevertheless, the low-GMI parameters for pairs of constellations $[\mX_\text{PSK},\mP_i]$ and $[\ntX_\text{PSK},\mU_3]$ are again equal. Particularly, $\alpha = 0.67$ for $[\mX_\text{PSK},\mP_2]$ and $\alpha = 0.76$ for $[\mX_\text{PSK},\mP_3]$.

The BICM-GMI for the two 8PSK constellations in Figs.~\ref{8PSK.Constellations.P2} and \ref{8PSK.Constellations.P3} is shown in Fig.~\ref{MI_transformed_PSK_constellations}. Also shown are the BICM-GMI and CM-MI for 8-PSK with uniform input distributions, for which $\alpha=0.62$ \cite[eq.~(56)]{Agrell10b} and $\alpha=\log_2 e$, resp., and the capacity of the AWGN channel. Again, the results show that the BICM-GMIs of the original constellation $[\mX_\text{PSK},\mP_i]$ and the transformed constellation $[\mathring{\mX}_\text{PSK},\mU_3]$ are in general different but converge in the low-SNR regime. The endpoints of the BICM-GMI curves are obtained from \eqref{min.EbN0}.

\begin{figure}
\newcommand{\scale}{0.9}
\psfrag{xlabel}[cc][cc][\scale]{$\Ebr/N_0$~[dB]}
\psfrag{ylabel}[bc][Bc][\scale]{$\Rc$~[bit/symbol]}
\psfrag{AWGNC}[cl][cl][\scale]{$\C^\text{AW}$}
\psfrag{CM-8PSKUnif}[cl][cl][\scale]{CM $[\mX_\text{PSK},\mU_3]$}
\psfrag{BICM-8PSKUnif}[cl][cl][\scale]{BICM $[\mX_\text{PSK},\mU_3]$}
\psfrag{BICM-8PSKS1}[cl][cl][\scale]{BICM $[\mX_\text{PSK},\mP_i]$}
\psfrag{BICM-8PSKT1}[cl][cl][\scale]{BICM $[\mathring{\mX}_\text{PSK},\mU_3]$}
\begin{center}
	\includegraphics[width=\columnwidth]{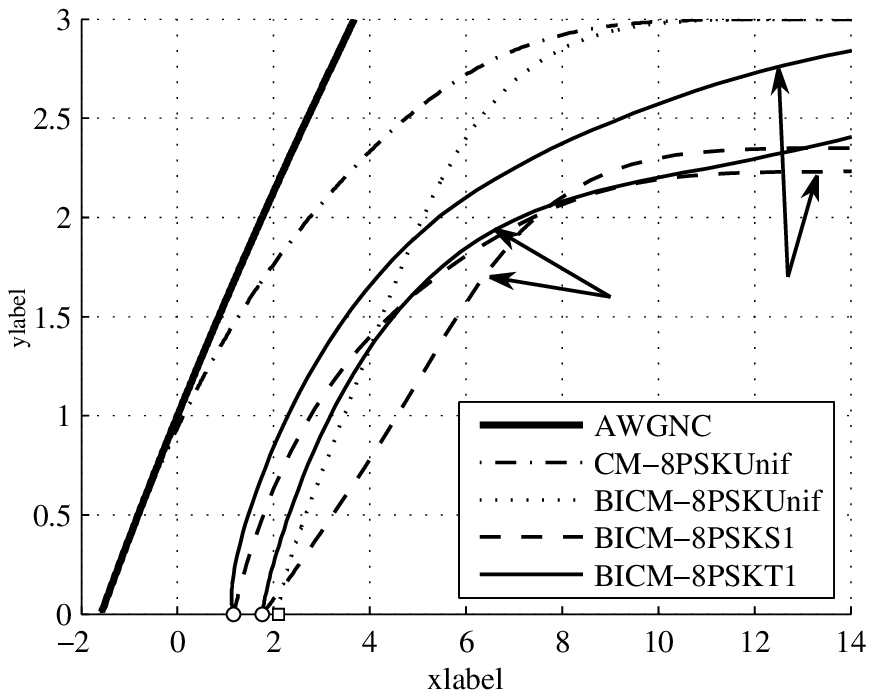}%
\overlay{-2.6cm}{3.9cm}{$\mP_2$}%
\overlay{-1.1cm}{3.9cm}{$\mP_3$}%
	\caption{BICM-GMI of the two probabilistically shaped 8-PSK constellations in Figs.~\ref{8PSK.Constellations.P2} and \ref{8PSK.Constellations.P3} and their transforms $[\ntX_\text{PSK},\mU_3]$. The CM-MI and BICM-GMI for uniform input distributions are also shown. The white circles and square show the endpoints $\alpha^{-1}$.}
    \label{MI_transformed_PSK_constellations}
\end{center}
\end{figure}
\end{example}

\begin{example}\label{Example.8QAM}
Consider the eight-level star-shaped QAM alphabet shown in Fig.~\ref{8QAM_constellations} (black circles), which we denote by $\mX_\text{8QAM}$. This alphabet is used with bit probabilities $\bb_4=[0.5,0.5,0.85]$, giving an input symbol probability $\mP_4$. In this figure we also show the transformed constellation $[\ntX_\text{8QAM},\mU_3]$, which according to Theorem~\ref{thB} has the same low-GMI parameters as $[\mX_\text{8QAM},\mP_4]$. This can be appreciated in Fig.~\ref{BICM_MI_8QAM_constellations}, where the corresponding BICM-GMIs are shown. Fig.~\ref{BICM_MI_8QAM_constellations} also shows how probabilistic shaping improves the BICM-GMI considerably over a wide range of SNRs.

\begin{figure}
\newcommand{\scale}{0.9}
\setlength{\xs}{1.6mm}
\setlength{\ys}{-2mm}
\psfrag{X0}[cl][cl][\scale]{}
\psfrag{X1}[cl][cl][\scale]{}
\psfrag{X2}[cl][cl][\scale]{}
\psfrag{X3}[cl][cl][\scale]{}
\psfrag{X4}[cl][cl][\scale]{}
\psfrag{X5}[cl][cl][\scale]{}
\psfrag{X6}[cl][cl][\scale]{}
\psfrag{X7}[cl][cl][\scale]{}
\psfrag{C0}[cl][cl][\scale]{\overlay{\xs}{\ys}{0}}
\psfrag{C1}[cl][cl][\scale]{\overlay{\xs}{\ys}{1}}
\psfrag{C2}[cl][cl][\scale]{\overlay{-\xs}{-\ys}{2}}
\psfrag{C3}[cl][cl][\scale]{\overlay{\xs}{-\ys}{3}}
\psfrag{C4}[cl][cl][\scale]{\overlay{1.3mm}{-5.5mm}{4}}
\psfrag{C5}[cl][cl][\scale]{\overlay{-5.8mm}{-3.5mm}{5}}
\psfrag{C6}[cl][cl][\scale]{\overlay{3mm}{1.5mm}{6}}
\psfrag{C7}[cl][cl][\scale]{\overlay{-3.8mm}{3.5mm}{7}}
\begin{center}
	\includegraphics{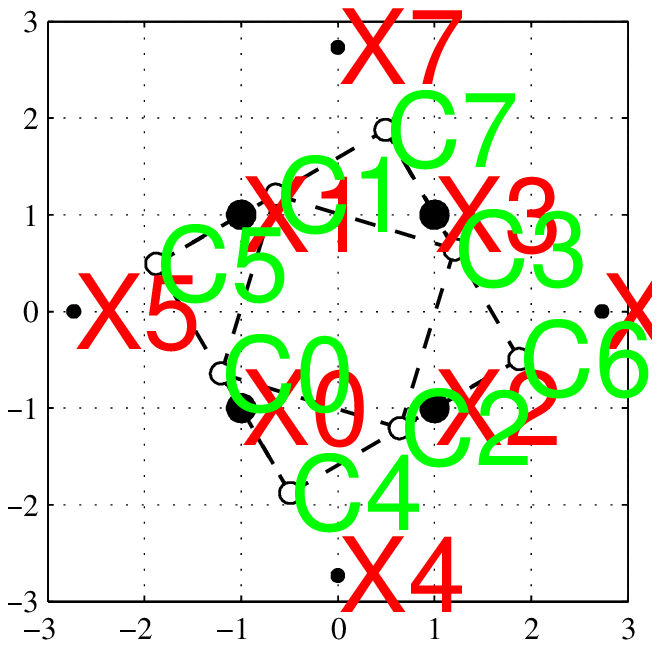}
	\caption{Star-shaped 8-QAM constellation $[\mX_\text{8QAM},\mP_4]$ with bit probabilities $\bb_4$ (black circles) and its transform $[\ntX_\text{8QAM},\mU_3]$ (white circles).}
	\label{8QAM_constellations}
\end{center}
\end{figure}

\begin{figure}
\newcommand{\scale}{0.9}
\psfrag{xlabel}[cc][cc][\scale]{$\Ebr/N_0$~[dB]}
\psfrag{ylabel}[bc][Bc][\scale]{$\Rc$~[bit/symbol]}
\psfrag{AWGNC}[cl][cl][\scale]{$\C^\text{AW}$}
\psfrag{BICM-GMI-Uniform}[cl][cl][\scale]{BICM $[\mX_\text{8QAM},\mU_3]$}
\psfrag{BICM-GMI-P}[cl][cl][\scale]{BICM $[\mX_\text{8QAM},\mP_4]$}
\psfrag{BICM-GMI-P-Trans}[cl][cl][\scale]{BICM $[\mathring{\mX}_\text{8QAM},\mU_3]$}
\begin{center}
	\includegraphics[width=\columnwidth]{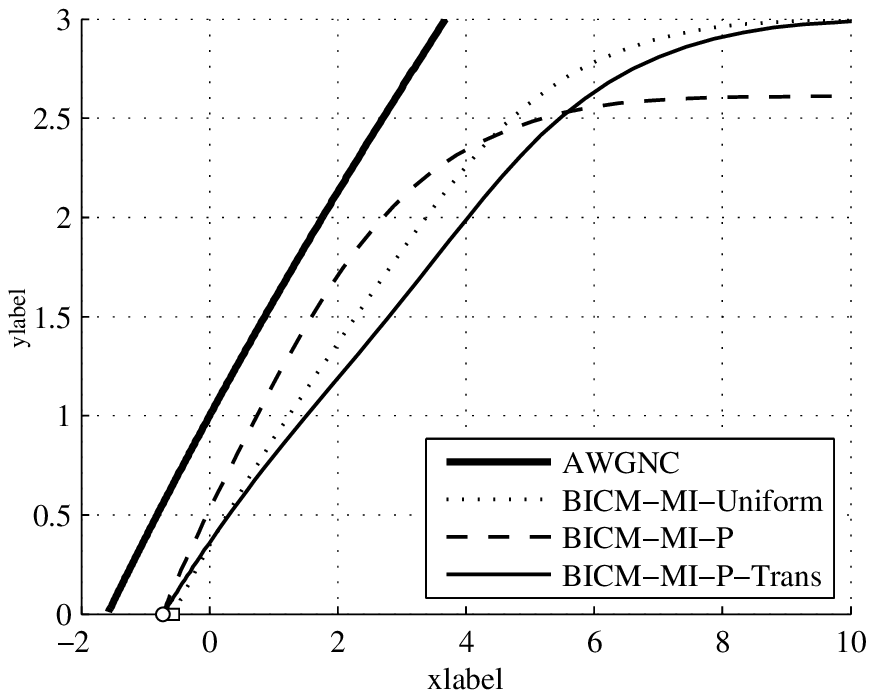}%
	\caption{BICM-GMI for $[\mX_\text{8QAM},\mU_3]$, $[\mX_\text{8QAM},\mP_4]$ and $[\mathring{\mX}_\text{8QAM},\mU_3]$. The AWGN capacity is also shown. The white circle and square show the endpoints $\alpha^{-1}$ with $\alpha=1.14$ and $1.18$.}
    \label{BICM_MI_8QAM_constellations}
\end{center}
\end{figure}
\end{example}

The results in Figs.~\ref{MI_transformed_QAM_constellations}, \ref{MI_transformed_PSK_constellations}, and \ref{BICM_MI_8QAM_constellations} also show other interesting properties of probabilistic shaping for BICM. In the high-SNR regime, the use of a nonuniform distribution results in a loss in GMI, i.e., the curves flatten out at a value below $m$~[bit/symbol], but for a wide range of moderately high SNR, the BICM-GMI is higher with probabilistic shaping. For the 16-QAM alphabet in Example~\ref{Example.16QAM-BRGC}, in the medium SNR regime, the use of nonequally likely symbols even gives a larger GMI than the CM-MI obtained with and a uniform distribution.

\section{First-order Optimal Constellations}\label{Sec:FOOBICM}

Having characterized the low-SNR behavior of the BICM-GMI of an arbitrary constellation, the next step is to search for optimal constellations in terms of the BICM-GMI at low SNR. The following definition formally defines BICM systems that achieve the SL.

\begin{definition}[FOO constellation]
The constellation $[\mX, \mP]$ is said to be first-order optimal (FOO) if a BICM system using $[\mX, \mP]$ achieves the SL $-1.59$~dB, i.e., $\alpha=\log_2\text{e}$.
\end{definition}

As discussed in Sec.~\ref{Sec:preliminaries:Shaping}, we have fixed the labeling to be the NBC, and thus, an FOO constellation is fully characterized by only two parameters, the input alphabet $\mX$ and its input distribution $\mP$, where the latter satisfies \eqref{pck-prod}. The analysis is nevertheless, without loss of generality, applicable to an arbitrary labeling by permuting the constellation, see Sec.~\ref{Sec:preliminaries:Shaping}.

\subsection{FOO Constellations for Uniform Distributions}\label{Sec:FOOBICM.UD}

In this section we review results on FOO constellations for BICM for uniform input distributions. The next theorem gives necessary and sufficient conditions for an input alphabet to be FOO if the binary labeling is the NBC and input distribution is uniform. 

\begin{theorem}\label{thC1}
The constellation $[\mX,\mU_m]$ is FOO if and only if \eqlab{thC1.1}{
\tbx_j = \bzero, \qquad\forall j\notin\set{1,2,4,\ldots,M/2}.
}
\end{theorem}

\begin{IEEEproof}
From \eqref{thC.2} and \eqref{thC.3}, $\alpha = \log_2\text{e}$ if and only if
\eq{
\sum_{i=0}^{M-1} \|\tbx_i\|^2 &= \sum_{k=0}^{m-1} \|\tbx_{2^k}\|^2
}
which gives \eqref{thC1.1}.
\end{IEEEproof}

This theorem was given in \cite[Th.~12]{Agrell10b}, where it was used to find FOO constellations for BICM when $\mP=\mU_m$. It offers an appealing intuitive geometrical interpretation: An input alphabet is FOO for a uniform input distribution if and only if it is a zero-mean linear projection of a hypercube. This behavior is illustrated in Example~\ref{ZeroMean.OTTO.Example} and also in \cite[Fig.~4]{Agrell10b}.

\subsection{FOO Constellations for Nonuniform Distributions}\label{Sec:FOOBICM.NUD}

In this section, we derive necessary and sufficient conditions for a BICM system, with an arbitrary input alphabet and probability distribution, to achieve the SL, i.e., we find FOO constellations for BICM. The conditions are derived by transforming an arbitrary constellation into another constellation with uniform probabilities using Theorem~\ref{thB} and applying Theorem~\ref{thC1} to this transformed constellation. Since Theorem~\ref{thC1} is expressed in terms of the HT, a relation between the HTs of $\mX$ and $\ntX$ is needed, which is illustrated by the bottom arrow in Fig.~\ref{visual_transform}. Such a relation is provided by Theorem~\ref{thD} and will be applied in the proofs of Theorems~\ref{thE} and \ref{thF}.

\begin{theorem}\label{thE}
The constellation $[\mX,\mP]$ is FOO if and only if the HT $\tX$ of $\mX$ satisfies both the following conditions:
\eqlab{thE1}{
\tbx_0 &= \sum_{k=0}^{m-1} \tbx_{2^k}\left(P_{C_k}(1)-P_{C_k}(0)\right) \\
\tbx_j &= \bzero, \qquad\forall j\notin\set{0} \cup \set{1,2,4,\ldots,M/2} \label{thE2}
.}
\end{theorem}

\begin{IEEEproof}
We will prove the theorem in two steps. First, we prove the ``if'' part by showing that \eqref{thE1} and \eqref{thE2} imply that $[\mX,\mP]$ is FOO. Second, we prove the ``only if'' part by showing that if $[\mX,\mP]$ is FOO, then \eqref{thE1} and \eqref{thE2} hold.

For the ``if'' part, suppose that \eqref{thE1} and \eqref{thE2} hold for a given constellation $[\mX,\mP]$. Applying \eqref{thE2} in \eqref{eq:thD1} yields for the HT $\tS$ of $\mS = \ntX$
\eqlab{E.3}{
\tbs_j &= \psi_j \bigg[ \tbx_0 \prod_{\substack{k=0\\n_{0,k} \ne n_{j,k}}}^{m-1}
\left(\pck(0)-\pck(n_{0,k})\right) &\nonumber\\
&\qquad + \sum_{l=0}^{m-1} \tbx_{2^l} \prod_{\substack{k=0\\n_{2^l,k} \ne n_{j,k}}}^{m-1}
\left(\pck(0)-\pck(n_{2^l,k})\right) \bigg], &\nonumber\\
&\hspace{5em} j=0,\ldots,M-1
.}
Since the bits $n_{0,k}=0$ for $k=0,\ldots,m-1$, the first product in \eqref{E.3} is always zero, except when $j=0$. (Recall that a product over $\varnothing$ in \eqref{eq:thD1} is defined as 1.) Furthermore, because of \eqref{nbc-poweroftwo}, the second product in \eqref{E.3} is zero whenever $j \notin \{0,2^l\}$ for some integer $l$. We can therefore identify three cases for \eqref{E.3}. First,
\eqlab{E.4}{
\tbs_j = \bzero, \quad j \notin \{0\} \cup \{1,2,\ldots,2^{m-1} \}.
}
Second, $j=0$ yields
\eqlab{E.5}{
\tbs_0 &= \psi_0 \left[ \tbx_0 + \sum_{l=0}^{m-1} \tbx_{2^l} \left( P_{C_l}(0) - P_{C_l}(n_{2^l,l}) \right)\right]
\nonumber\\
&= \bzero
}
because of \eqref{thE1}. And third, letting $j=2^i$ for an integer $i$,
\eqlab{E.6}{
\tbs_{2^i} = \psi_{2^i} \sum_{l=0}^{m-1} \tbx_{2^l} \prod_{\substack{k=0\\n_{2^i,k} \ne n_{2^l,k}}}^{m-1}
  \left( P_{C_k}(0) - P_{C_k}(n_{2^l,k}) \right), \nonumber\\
  i = 0,\ldots,m-1
.}
When $l\ne i$, the product in \eqref{E.6} includes two factors, $k=l$ and $k=i$. For $k=i$, $n_{2^l,k} = 0$ and the whole product is $0$. Therefore, only $l=i$ contributes to the sum in \eqref{E.6}. When $l=i$, the product in \eqref{E.6} is again over $\varnothing$ and \eqref{E.6} becomes
\eqlab{E.7}{
  \tbs_{2^i} = \psi_{2^i} \tbx_{2^i}
.}

Combining the three cases \eqref{E.4}, \eqref{E.5}, and \eqref{E.7} yields
\eq{
\tbs_j = \begin{cases}
\psi_j \tbx_j, & j \in \{1,2,\ldots,2^{m-1} \}, \\
\bzero, & \text{otherwise}.
\end{cases}
}
Using Theorem~\ref{thC1}, we conclude that the constellation $[\ntX,\mU_m]$ is FOO. Finally, Theorem~\ref{thB} implies that $[\mX,\mP]$ is also FOO, which completes the proof of the ``if'' part.

For the ``only if'' part, assume that $[\mX,\mP]$ is FOO. By Theorem~\ref{thB}, $[\ntX,\mU_m]$ is also FOO, and by Theorem~\ref{thC1}, $\tbs_i = \bzero$ for any $i$ that is not a power of two, where $\tS$ is the HT of $\mS = \ntX$. We will now use Theorem~\ref{thD} to translate the condition on $\tS$ into conditions on $\tX$.

If $\tbs_i = \bzero$ for $i \notin \{1,2,\ldots,2^{m-1}\}$, then the summation over $i=0,\ldots,M-1$ in \eqref{eq:thD2} can be reduced to a summation over $i = 2^l$ for $l=0,\ldots,m-1$,
\eqlab{E.8}{
\tbx_j &= \sum_{l=0}^{m-1} \frac{\tbs_{2^l}}{\psi_{2^l}}
\prod_{\substack{k=0\\n_{j,k} \ne n_{2^l,k}}}^{m-1}
\left( \pck(n_{2^l,k})-\pck(0) \right), \nonumber\\
&\hspace{10em} j=0,\ldots,M-1
.}
Due to \eqref{nbc-poweroftwo}, the product in \eqref{E.8} is nonzero only if the product is over $k\in \varnothing$ or over the single-element set $k \in \{l\}$, i.e., if $j=0$ or $j=2^l$ for some integer $l$, resp. Again, three cases can be identified. First, if $j \notin \{0 \} \cup \{1,2,\ldots,2^{m-1}\}$, then the product in \eqref{E.8} includes at least one $k\ne l$ for every $l$ and
\eqlab{E.9}{
\tbx_j = \bzero,\qquad j \notin \{0 \} \cup \{1,2,\ldots,2^{m-1}\}
.}
Second, for $j=0$, the product comprises only one factor, $k=l$, and
\eqlab{E.10}{
\tbx_0 &= \sum_{l=0}^{m-1} \frac{\tbs_{2^l}}{\psi_{2^l}} \left( P_{C_l}(1) - P_{C_l}(0) \right)
.}
And third, setting $j=2^i$,
\eqlab{E.11}{
\tbx_{2^i} = \sum_{l=0}^{m-1} \frac{\tbs_{2^l}}{\psi_{2^l} } \prod_{\substack{k=0\\n_{2^i,k} \ne n_{2^l,k}}}^{m-1}
  \left( P_{C_k}(n_{2^l,k}) - P_{C_k}(0) \right), \nonumber\\
  i = 0,\ldots,m-1
.}
As explained after \eqref{E.6}, the product in \eqref{E.11} is $1$ if $l=i$ (product over $\varnothing$) and $0$ otherwise. Thus, the summation in \eqref{E.11} can be reduced to just one term, $l=i$, and
\eqlab{E.12}{
\tbx_{2^i} = \frac{\tbs_{2^i}}{\psi_{2^i}}, \qquad i=0,\ldots, m-1
.}

Combining the three cases \eqref{E.9}, \eqref{E.10}, and \eqref{E.12} yields
\eq{
\tbx_j = \begin{cases}
\sum_{l=0}^{m-1} \tbs_{2^l} \frac{P_{C_l}(1)-P_{C_l}(0)}{\psi_{2^l}}, & j=0, \\
\frac{\tbs_j}{\psi_j}, & j \in \{1,2,\ldots,2^{m-1} \} \\
\bzero, & \text{otherwise}
\end{cases}
}
which satisfies \eqref{thE1} and \eqref{thE2}. This completes the proof of the ``only if'' part.
\end{IEEEproof}

\begin{remark}
Only \eqref{thE1} depends on the input distribution, not \eqref{thE2}. In view of Theorem~\ref{thC1}, the only difference between FOO constellations with uniform and nonuniform distributions lies in $\tbx_0$. The final theorem gives this statement a more intuitive interpretation.
\end{remark}

\begin{theorem}\label{thF}
The constellation $[\mX,\mP]$ is FOO if and only if both the following conditions hold:
\eqlab{thF1}{
\bmu &= \bzero \\
\tbx_j &= \bzero, \qquad\forall j\notin\set{0} \cup \set{1,2,4,\ldots,M/2}. \label{thF2}
}
\end{theorem}

\begin{IEEEproof}
We wish to prove that if \eqref{thE2} (or equivalently \eqref{thF2}) holds, then \eqref{thE1} and \eqref{thF1} are equivalent.

For any constellation $[\mX,\mP]$, the mean $\bmu = \tbs_0$, where $\tS$ is the HT of $\mS = \ntX$. This follows from Theorem~\ref{thB} and \eqref{thC.1}. The mean $\bmu$ can be calculated by letting $i=0$ in \eqref{eq:thD1} and \eqref{psi} as
\eqlab{F.3}{
\bmu = \psi_0 \sumi \tbx_i \prod_{\substack{k=0\\n_{i,k} = 1}}^{m-1} \left( \pck(0)-\pck(1) \right)
}
where $\psi_0 = 1$.

In this theorem, we are only interested in constellations that satisfy \eqref{thE2} or equivalently \eqref{thF2}. For such constellations, the sum in \eqref{F.3} includes at most $m+1$ nonzero terms, namely,
\eq{
\bmu = \tbx_0 + \sum_{l=0}^{m-1} \tbx_{2^l} \left( P_{C_l}(0) - P_{C_l}(1) \right)
.}
This expression makes \eqref{thE1} and \eqref{thF1} equivalent.
\end{IEEEproof}

Based on Theorem~\ref{thC1}, the result in Theorem~\ref{thF} can be understood as follows. If a constellation with a uniform input distribution is FOO, it will still be FOO for any other input distribution $\bb$ provided that the input alphabet is translated to be zero mean. In view of the geometrical interpretation of Theorem~\ref{thC1} given in \cite[Th.~12]{Agrell10b}, the result in Theorem~\ref{thF} also states that a constellation is FOO if and only if its input alphabet is a linear projection of a hypercube \emph{and} it has zero mean.
 
We also note that the zero-mean condition in Theorem~\ref{thF} is the same that guarantees FOO for the CM-MI \cite[Footnote~12]{Agrell10b}\footnote{The parameter $\alpha$ for the CM-MI is \cite[Th.~7]{Agrell10b} $\alpha= \log_2\text{e}(1-{\|\bmu\|^2}/{\Es})$.}. This implies that the only difference between FOO constellations for the CM-MI and the BICM-GMI lies on the extra constraint on the input alphabet to be a linear projection of a hypercube.

\subsection{Numerical Examples}\label{Sec:FOOBICM.foo-examples}

In this subsection we give numerical examples to illustrate the analytical results presented in this paper.

\begin{figure}
\newcommand{\scale}{0.85}
\psfrag{x0}[cc][cc][\scale]{$0$}
\psfrag{x1}[cc][cc][\scale]{$1$}
\psfrag{x2}[cc][cc][\scale]{$2$}
\psfrag{x3}[cc][cc][\scale]{$3$}
\psfrag{x4}[cc][cc][\scale]{$4$}
\psfrag{x5}[cc][cc][\scale]{$5$}
\psfrag{x6}[cc][cc][\scale]{$6$}
\psfrag{x7}[cc][cc][\scale]{$7$}
\begin{center}
	\includegraphics{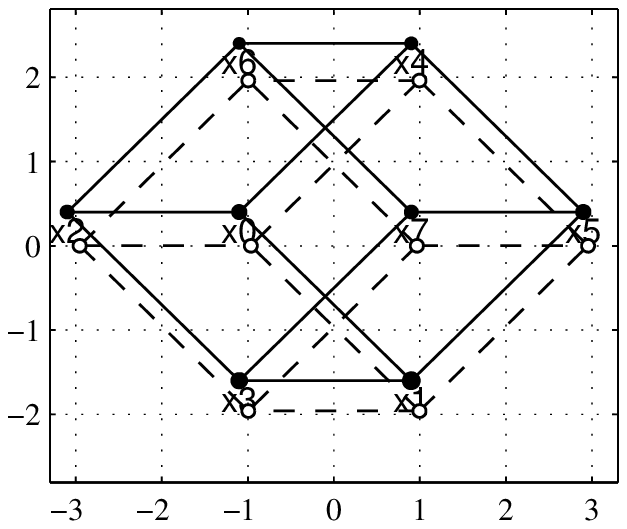}
	\caption{The FOO constellations $[\mX_{\text{8AMPM}}',\mP_5]$ (black circles) and $[\ntX_{\text{8AMPM}}',\mU_3]$ (white circles).}
    \label{shaped_OTTO_P6}
\end{center}
\end{figure}

\begin{example}\label{ZeroMean.OTTO.Example}
Consider the so-called 8-AMPM alphabet \cite{Ungerboeck82}
\begin{align}\label{ZeroMean.OTTO.Example.X}
\mX_{\text{8AMPM}}=
\begin{bmatrix}
-1&1&-3&-1&1&3&-1&1\\
0&-2&0&-2&2&0&2&0 
\end{bmatrix}^\T
\end{align}
which corresponds to a projected hypercube. The constellation $[\mX_{\text{8AMPM}},\mP]$ was shown to be FOO for $\mP=\mU_3$ in \cite[Example~4]{Agrell10b}. In view of Theorem~\ref{thF}, it is FOO for any $\mP$ if it has zero mean. Using \eqref{ZeroMean.OTTO.Example.X} and \eqref{pck-prod} in \eqref{thA1}, we find (after some algebra) that
\eqlab{exOTTO.bmu}{
\bmu = \begin{bmatrix}
1+2\left( P_{C_1}(0)-P_{C_0}(0)-P_{C_2}(0) \right) \\
2\left( P_{C_0}(0)-P_{C_2}(0) \right)
\end{bmatrix}^\T
.}
For example, the bit probabilities $\bb_5=[0.40,0.55,0.60]$ give an input distribution $\mP_5$ for which the mean \eqref{exOTTO.bmu} is $\bmu=[0.10,-0.40]$. We define another alphabet $\mX_{\text{8AMPM}}'$ by subtracting $\bmu$ from each element in $\mX_{\text{8AMPM}}$. Fig.~\ref{shaped_OTTO_P6} shows the translated constellation $[\mX_{\text{8AMPM}}',\mP_5]$ along with $[\ntX_{\text{8AMPM}}',\mU_3]$, where $\ntX_{\text{8AMPM}}'$ is the transform of $\mX_{\text{8AMPM}}'$ for the distribution $\mP_5$. They are both zero-mean projected hypercubes and thus FOO according to Theorem~\ref{thF}. This can be observed from the BICM-GMI curve for $[\mX_{\text{8AMPM}}',\mP_5]$ in Fig.~\ref{MI_FOO_constellations}.

\begin{figure}
\newcommand{\scale}{0.85}
\psfrag{x0}[cc][cc][\scale]{$0$}
\psfrag{x1}[cc][cc][\scale]{$1$}
\psfrag{x2}[cc][cc][\scale]{$2$}
\psfrag{x3}[cc][cc][\scale]{$3$}
\psfrag{x4}[cc][cc][\scale]{$4$}
\psfrag{x5}[cc][cc][\scale]{$5$}
\psfrag{x6}[cc][cc][\scale]{$6$}
\psfrag{x7}[cc][cc][\scale]{$7$}
\begin{center}
	\includegraphics{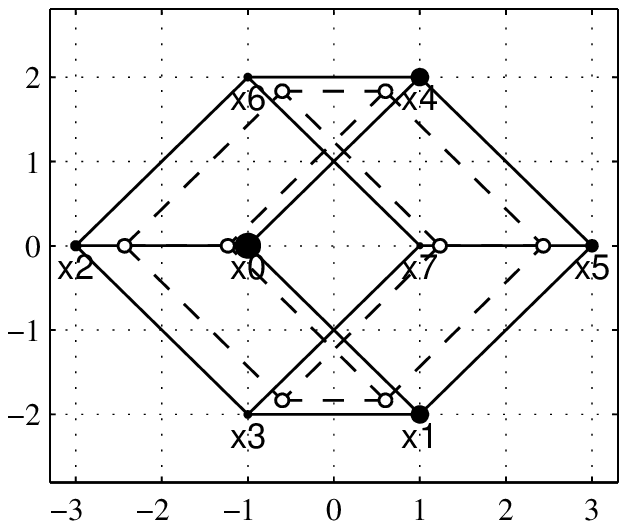}
	\caption{The constellations $[\mX_{\text{8AMPM}},\mP_6]$ (black circles) and $[\ntX_\text{8AMPM},\mU_3]$ (white circles). They both have zero mean and look like cubes, which indicates that they are FOO.}
    \label{shaped_OTTO_P7}
\end{center}
\end{figure}

The exemplified method holds in full generality: Any alphabet that is FOO with a uniform input distribution is FOO also with an arbitrary nonuniform distribution, if it is translated to zero mean. Furthermore, \emph{all} nonuniform FOO constellations can be constructed in this manner.

For certain distributions, the mean \eqref{exOTTO.bmu} is zero without translation. Specifically, $\bmu=\bzero$ if and only if
\begin{align}
\label{ZeroMean.OTTO.Example.Pcond}
P_{C_0}(0)=P_{C_2}(0)=P_{C_1}(0)/2+1/4.
\end{align}
Clearly, the uniform case ($P_{C_0}(0)=P_{C_1}(0)=P_{C_2}(0)=1/2$) analyzed in \cite{Agrell10b} fulfills \eqref{ZeroMean.OTTO.Example.Pcond}. More interestingly, when any other vector of bit probabilities fulfilling \eqref{ZeroMean.OTTO.Example.Pcond} is used, the resulting constellation will be FOO. This is the case for instance with $\bb_6=[0.70,0.90,0.70]$. The obtained constellation, denoted by $[\mX_{\text{8AMPM}},\mP_6]$ is illustrated in Fig.~\ref{shaped_OTTO_P7} along with its transform $[\ntX_\text{8AMPM},\mU_3]$. Graphically, both alphabets look like cubes, although viewed from different angles, which is precisely what Theorems~\ref{thC1} and \ref{thF} predict.

In Fig.~\ref{MI_FOO_constellations}, we show the BICM-GMI for the zero-mean constellations $[\mX_{\text{8AMPM}}',\mP_5]$ and $[\mX_{\text{8AMPM}},\mP_6]$ as well as for the constellations $[\ntX_{\text{8AMPM}}',\mU_3]$ and $[\ntX_{\text{8AMPM}},\mU_3]$. As expected, all the GMIs converge at the SL for low SNR. For these two cases, the transformed alphabets with uniform input distributions give larger GMI for all SNRs compared to the corresponding nonuniform ones. This is however not always the case, cf.~Fig.~\ref{MI_transformed_PSK_constellations} with $\mP_2$.

\begin{figure}
\newcommand{\scale}{0.9}
\psfrag{xlabel}[cc][cc][\scale]{$\Ebr/N_0$~[dB]}
\psfrag{ylabel}[bc][Bc][\scale]{$\Rc$~[bit/symbol]}
\psfrag{AWGNC}[cl][cl][\scale]{$\C^\text{AW}$}
\psfrag{BI-OTTOP6}[cl][cl][\scale]{$[\mX_{\text{8AMPM}}',\mP_5]$}
\psfrag{BI-OTTOP6T}[cl][cl][\scale]{$[\ntX_{\text{8AMPM}}',\mU_3]$}
\psfrag{BI-OTTOP7}[cl][cl][\scale]{$[\mX_{\text{8AMPM}},\mP_6]$}
\psfrag{BI-OTTOP7T}[cl][cl][\scale]{$[\ntX_{\text{8AMPM}},\mU_3]$}
\begin{center}
	\includegraphics[width=\columnwidth]{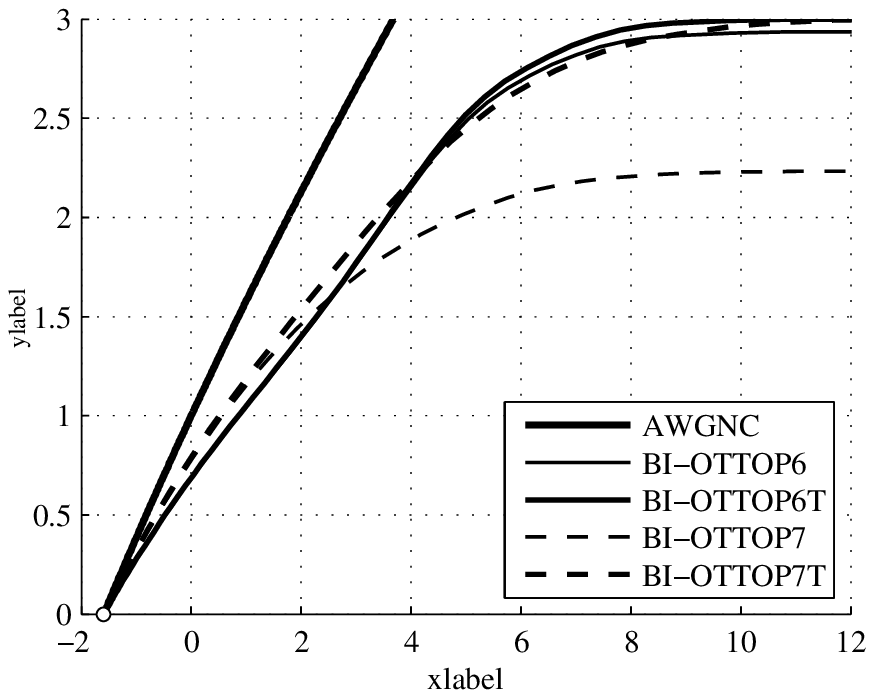}
	\caption{BICM-GMI for the four FOO constellations in Figs.~\ref{shaped_OTTO_P6} and \ref{shaped_OTTO_P7}. The SL is shown with a white circle.}
    \label{MI_FOO_constellations}
\end{center}
\end{figure}

\end{example}

\begin{example}\label{8PAM.Example}
$M$-PAM alphabets have been shown to be FOO if the NBC is used with a uniform input distribution, i.e, the constellation $[\mX_{\text{PAM}},\mU_m]$ with $\mX_{\text{PAM}}=[-(M-1),-(M-3),\ld,M-1]$ is FOO \cite{Stierstorfer09a} \cite[Th.~14]{Agrell10b}. In this example, we study the first-order behavior of the 8-PAM alphabet $\mX_{\text{PAM}}'=[-7,7,-1,1,-5,5,-3,3]$, where the order of the points represents the BRGC. The constellation $[\mX_{\text{PAM}}',\mP]$ is known not to be FOO for $\mU_3$ \cite[Th.~3]{Martinez08b}.

The BICM-GMI of $\mX_{\text{PAM}}'$ is shown in Fig.~\ref{MI_8PAM_constellations} for the set of bit probabilities $\bb_7=[0.5,p,p]$ for different values of $p$. For $p=0.5$, the uniform distribution is obtained. As $p$ decreases, the Gray-labeled constellation approaches a zero-mean binary alphabet, which is FOO \cite[Fig.~2]{Fabregas10a} \cite[Fig.~3~(b)]{Agrell10b}. Fig.~\ref{MI_8PAM_constellations} illustrates the tradeoff between the low- and high-SNR regimes: The SL can be approached by decreasing $p$, but this causes a decrease in GMI in the high-SNR regime. Alternatively, the SL can be attained by switching from the BRGC to the NBC, but this also comes with a heavy penalty at higher SNRs.

\begin{figure}
\newcommand{\scale}{0.9}
\psfrag{xlabel}[cc][cc][\scale]{$\Ebr/N_0$~[dB]}
\psfrag{ylabel}[bc][Bc][\scale]{$\Rc$~[bit/symbol]}
\psfrag{AWGNC}[cl][cl][\scale]{$\C^\text{AW}$}
\psfrag{BI8PAMU}[cl][cl][\scale]{$[\mX_{\text{PAM}}',\mU_3]$}
\psfrag{BI8PAM03}[cl][cl][\scale]{$p=0.3$}
\psfrag{BI8PAM02}[cl][cl][\scale]{$p=0.2$}
\psfrag{BI8PAM01}[cl][cl][\scale]{$p=0.1$}
\psfrag{BI8PAM005}[cl][cl][\scale]{$p=0.05$}
\psfrag{BI8PAM001}[cl][cl][\scale]{$p=0.01$}
\psfrag{BI8PAMUp}[cl][cl][\scale]{$[\mX_{\text{PAM}},\mU_3]$}
\psfrag{x0}[bc][bc][\scale]{$x_0'$}
\psfrag{x1}[bc][bc][\scale]{$x_4'$}
\psfrag{x2}[bc][bc][\scale]{$x_6'$}
\psfrag{x3}[bc][bc][\scale]{$x_2'$}
\psfrag{x4}[bc][bc][\scale]{$x_3'$}
\psfrag{x5}[bc][bc][\scale]{$x_7'$}
\psfrag{x6}[bc][bc][\scale]{$x_5'$}
\psfrag{x7}[bc][bc][\scale]{$x_1'$}
\psfrag{l0}[tc][tc][\scale]{$000$}
\psfrag{l1}[tc][tc][\scale]{$\,001$}
\psfrag{l2}[tc][tc][\scale]{$011$}
\psfrag{l3}[tc][tc][\scale]{$010$}
\psfrag{l4}[tc][tc][\scale]{$110$}
\psfrag{l5}[tc][tc][\scale]{$111$}
\psfrag{l6}[tc][tc][\scale]{$101$}
\psfrag{l7}[tc][tc][\scale]{$100$}
\begin{center}
	\includegraphics[width=\columnwidth]{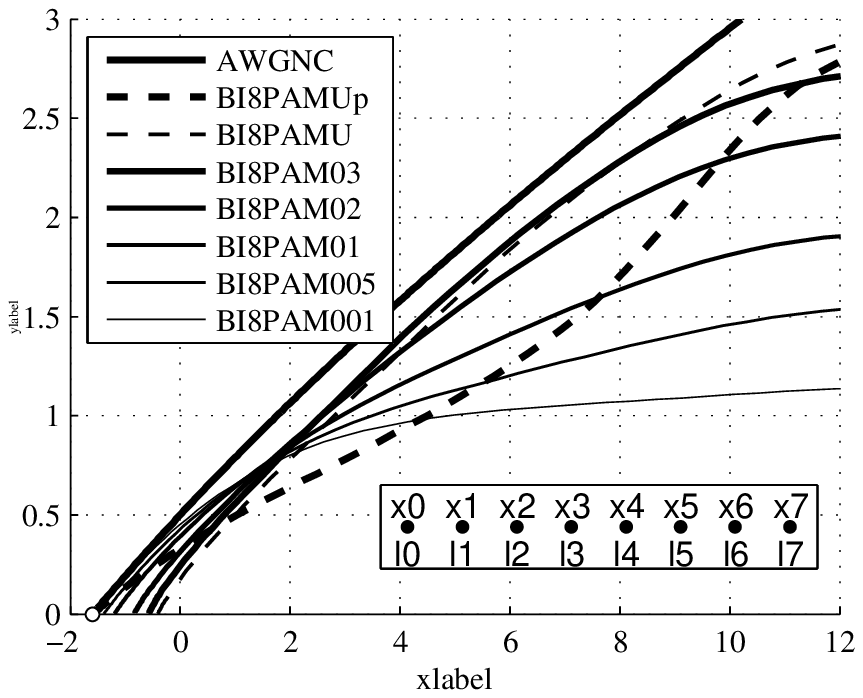}
	\caption{The BICM-GMI for the 8-PAM alphabet $\mX_{\text{PAM}}'$ labeled by the BRGC with uniform input distribution and bit probabilities $\bb_7=[0.5,p,p]$ for different values of $p$. The constellations approach a binary FOO constellation as $p \rightarrow 0$. The NBC-labeled 8-PAM alphabet $\mX_{\text{PAM}}$, although FOO with a uniform distribution, is considerably weaker than $\mX_{\text{PAM}}'$ for a wide range of SNRs.}
    \label{MI_8PAM_constellations}
\end{center}
\end{figure}

\end{example}

\section{Conclusions}\label{Sec:Conclusions}

There exists a closed-form mapping between any probabilistically shaped constellation and a constellation with uniform input distribution, such that the two systems have the same low-SNR first-order behavior (Definition~\ref{new-transform} and Theorem~\ref{thB}). Thus, the combination of probabilistic and geometric shaping is equivalent to pure geometric shaping at low SNR.

We are particularly interested in BICM systems that attain the SL $-1.59$ dB at asymptotically low SNR, i.e., in FOO constellations. Somewhat disappointingly, the set of probabilistically shaped FOO constellation is no larger than the set of FOO constellations with uniform distributions, disregarding translations of the whole input alphabet. Both sets can be fully characterized as the set of linear projections of a hypercube, translated to have zero mean for the considered input distribution (Theorems \ref{thC1} and \ref{thE}; cf.~Figs.~\ref{shaped_OTTO_P6} and \ref{shaped_OTTO_P7}). Although non-FOO constellations for BICM can be improved by probabilistic shaping (Fig.~\ref{MI_8PAM_constellations}), it is impossible to make them FOO except in degenerate cases (by setting some probabilities equal to zero).

\appendix[Properties of the Transform]

In this Appendix, some theoretical properties of the new transform defined in Section~\ref{Sec:preliminaries:new-transform} are proved. In the first section, the ``sum-product lemma'' is established, which will be used extensively throughout the appendix. In the following two sections, Lemma~\ref{lemmaG} and Theorem~\ref{thD} are proved.

\subsection{The Sum-Product Lemma}
Many properties of NBC-labeled constellations and their transforms can be expressed as the sum of products of certain functions, where each function depends on one bit position. Such expressions, which occur frequently in the two subsequent sections, can be resolved using this general lemma.

\begin{lemma}\label{lemmaH}
Let $f_{k,u}$ for $k=0,\ldots,m-1$ and $u\in \mcB$ be any real numbers. Then
\eq{
\sum_{i=0}^{M-1} \prod_{k=0}^{m-1} f_{k,n_{i,k}} = \prod_{k=0}^{m-1} (f_{k,0}+f_{k,1})
.}
\end{lemma}

{
\renewcommand{\IEEEQED}{\raisebox{4ex}[0ex][0ex]{\IEEEQEDopen}\vspace*{-1ex}}
\begin{IEEEproof}
A summation over $i = 0,\ldots, 2^m-1$ is equivalent to $m$ sums over $i_k \in \mcB$, where $k=0,\ldots,m-1$ and $i=i_0+2i_1+\cdots+2^{m-1}i_{m-1}$. With this notation, $n_{i,k} = i_k$ and
\eq{
\sum_{i=0}^{M-1} \prod_{k=0}^{m-1} f_{k,n_{i,k}}
&= \sum_{i_0\in\mcB} \sum_{i_1\in\mcB} \cdots\!\! \sum_{i_{m-1}\in\mcB} \prod_{k=0}^{m-1} f_{k,i_k}\\
&= \sum_{i_0\in\mcB} \sum_{i_1\in\mcB} \cdots\!\! \sum_{i_{m-1}\in\mcB} f_{0,i_0}f_{1,i_1}\cdots f_{m-1,i_{m-1}} \!\\
&= \sum_{i_0\in\mcB} f_{0,i_0} \sum_{i_1\in\mcB} f_{1,i_1} \cdots\!\! \sum_{i_{m-1}\in\mcB} f_{m-1,i_{m-1}}\\
&= \prod_{k=0}^{m-1} \sum_{i_k\in\mcB} f_{k,i_k}
.}
\end{IEEEproof}
}

\subsection{The Transform Coefficients}
Lemma~\ref{lemmaG}, which lists two fundamental properties of the transform coefficients $\gamma_{i,j}$, was given in Section~\ref{Sec:preliminaries:new-transform}.

\begin{IEEEproof}[Proof of Lemma~\ref{lemmaG}]
The two parts of Lemma~\ref{lemmaG} will now be proved separately. First, the definition of $\gamma_{i,j}$ in \eqref{gammadef} yields

\eq{
&\sum_{i=0}^{M-1} \gamma_{i,l} \gamma_{i,j}=\sum_{i=0}^{M-1} \prod_{k=0}^{m-1}
	\Big[(-1)^{\bar{n}_{i,k}n_{l,k}}\sqrt{P_{C_k}(0)}\\
	&\qquad\quad+(-1)^{n_{i,k}\bar{n}_{l,k}}\sqrt{P_{C_k}(1)}\Big] \\
	&\qquad\cdot \Big[(-1)^{\bar{n}_{i,k}n_{j,k}}\sqrt{P_{C_k}(0)} \\
	&\qquad\quad+(-1)^{n_{i,k}\bar{n}_{j,k}}\sqrt{P_{C_k}(1)}\Big] \\
&=\sum_{i=0}^{M-1} \prod_{k=0}^{m-1}
	\Big[(-1)^{\bar{n}_{i,k}(n_{l,k}+n_{j,k})}P_{C_k}(0) \\
	&\qquad+(-1)^{\bar{n}_{i,k}n_{l,k}+n_{i,k}\bar{n}_{j,k}}\sqrt{P_{C_k}(0)P_{C_k}(1)} \\
	&\qquad+(-1)^{n_{i,k}\bar{n}_{l,k}+\bar{n}_{i,k}n_{j,k}}\sqrt{P_{C_k}(0)P_{C_k}(1)} \\
	&\qquad+(-1)^{n_{i,k}(\bar{n}_{l,k}+\bar{n}_{j,k})}P_{C_k}(1) \Big] \\
&=\sum_{i=0}^{M-1} \prod_{k=0}^{m-1}(-1)^{n_{i,k}(n_{l,k}+n_{j,k})} \\
	&\qquad\cdot\Big[(-1)^{n_{l,k}+n_{j,k}}P_{C_k}(0)+P_{C_k}(1) \\
	&\qquad\quad+(-1)^{n_{i,k}}\left[ (-1)^{n_{l,k}}+(-1)^{n_{j,k}} \right]\sqrt{P_{C_k}(0)P_{C_k}(1)} \Big]
}
where the last equality follows by repeatedly using the identities $\bar{u} = 1-u$ and $(-1)^u = (-1)^{-u}$ for $u \in \mcB$. Lemma~\ref{lemmaH} now yields
\eqlab{G2a}{
&\sum_{i=0}^{M-1} \gamma_{i,l} \gamma_{i,j}=
\prod_{k=0}^{m-1}\Big[
	(-1)^{n_{l,k}+n_{j,k}}P_{C_k}(0)+P_{C_k}(1) \nonumber\\
	&\qquad+\left[ (-1)^{n_{l,k}}+(-1)^{n_{j,k}} \right]\sqrt{P_{C_k}(0)P_{C_k}(1)} \nonumber\\
	&\qquad+(-1)^{n_{l,k}+n_{j,k}}\Big( (-1)^{n_{l,k}+n_{j,k}}P_{C_k}(0)+P_{C_k}(1) \nonumber\\
	&\qquad\quad-\left[ (-1)^{n_{l,k}}+(-1)^{n_{j,k}} \right]\sqrt{P_{C_k}(0)P_{C_k}(1)} \Big)\Big]
	\nonumber\\
&=\prod_{k=0}^{m-1}\Big[
	\left( (-1)^{n_{l,k}+n_{j,k}}+1 \right) P_{C_k}(0) \nonumber\\
	&\qquad+\left( 1+(-1)^{n_{l,k}+n_{j,k}} \right) P_{C_k}(1) \nonumber\\
	&\qquad+\left[ (-1)^{n_{l,k}}+(-1)^{n_{j,k}}-(-1)^{n_{l,k}}-(-1)^{n_{j,k}} \right] \nonumber\\
	&\qquad\quad \cdot\sqrt{P_{C_k}(0)P_{C_k}(1)} \Big] \nonumber\\
&=\prod_{k=0}^{m-1}\left( 1+(-1)^{n_{l,k}+n_{j,k}} \right)
}
where the last step follows because $P_{C_k}(0)+P_{C_k}(1)=1$. The factors in \eqref{G2a} are either $2$ or $0$, depending on whether $n_{l,k}=n_{j,k}$ or $n_{l,k}\ne n_{j,k}$ for the particular bit position $k$. Thus, $\sumi \gamma_{i,l} \gamma_{i,j}$ is either $2^m=M$ or $0$, depending on whether $l$ and $j$ have \emph{all} bits equal or not. This completes the proof of \eqref{G2}.

To prove the second part of Lemma~\ref{lemmaG}, which is \eqref{G4}, we observe from \eqref{hcoeff} and \eqref{gammadef} that
\eqlab{G4a}{
\sumi h_{l,i} \gamma_{i,j}
&= \sumi \prodk (-1)^{n_{l,k}n_{i,k}} \nonumber\\
&\hspace{-4em}\cdot \prodk \left[ (-1)^{\bar{n}_{i,k}n_{j,k}}\sqrt{\pck(0)}
+ (-1)^{n_{i,k}\bar{n}_{j,k}}\sqrt{\pck(1)} \right] \nonumber\\
&= \sumi \prodk \phi_{k,n_{i,k}}
}
where
\eq{
\phi_{k,u} &\triangleq
(-1)^{un_{l,k}+\bar{u}n_{j,k}}\sqrt{\pck(0)} \nonumber\\
&\qquad+ (-1)^{u n_{l,k}+u\bar{n}_{j,k}}\sqrt{\pck(1)}
.}
Intending to apply Lemma~\ref{lemmaH} to \eqref{G4a}, we first calculate the quantity
\eq{
\phi_{k,0} + \phi_{k,1}
&= (-1)^{n_{j,k}}\sqrt{\pck(0)} + \sqrt{\pck(1)} \nonumber\\
&\quad +(-1)^{n_{l,k}}\sqrt{\pck(0)} + (-1)^{n_{l,k}+\bar{n}_{j,k}} \sqrt{\pck(1)}
}
for $k=0,\ldots,m-1$. For reasons that will soon become clear, we extract a common factor $(-1)^{n_{j,k}n_{l,k}}$ from all terms, obtaining
\eq{
\phi_{k,0} + \phi_{k,1}
&= (-1)^{n_{j,k}n_{l,k}} \nonumber\\
&\quad\cdot \Big[ \left((-1)^{n_{j,k}\bar{n}_{l,k}}+(-1)^{\bar{n}_{j,k}n_{l,k}}\right)\sqrt{\pck(0)} \nonumber\\
&\quad\quad
+\left((-1)^{n_{j,k}n_{l,k}}+(-1)^{\bar{n}_{j,k}\bar{n}_{l,k}}\right)\sqrt{\pck(1)} \Big]
.}
The coefficient in front of $\sqrt{\pck(0)}$ is $2$ if $n_{j,k} = n_{l,k}$ and $0$ otherwise. Similarly, the coefficient in front of $\sqrt{\pck(1)}$ is $0$ if $n_{j,k} = n_{l,k}$ and $2$ otherwise. Thus, for every $k$, the expression depends on either $\pck(0)$ or $\pck(1)$ but not both, namely,
\eqlab{G4b}{
\phi_{k,0} + \phi_{k,1}
&= \begin{cases}
(-1)^{n_{j,k}n_{l,k}}\cdot 2\sqrt{\pck(0)}, & n_{j,k} = n_{l,k} \\
(-1)^{n_{j,k}n_{l,k}}\cdot 2\sqrt{\pck(1)}, & n_{j,k} \ne n_{l,k}
\end{cases} \nonumber\\
&= 2(-1)^{n_{j,k}n_{l,k}}\sqrt{\pck(n_{j,k} \oplus n_{l,k})} \nonumber\\
&= 2(-1)^{n_{j,k}n_{l,k}}\sqrt{\pck(n_{j \oplus l,k})}
}
according to \eqref{nbc-oplus}.

Using \eqref{G4b}, we are now ready to apply Lemma~\ref{lemmaH} to \eqref{G4a}, which yields
\eq{
\sumi h_{i,l} \gamma_{i,j} &= \prodk 2(-1)^{n_{j,k}n_{l,k}}\sqrt{\pck(n_{j \oplus l,k})}
.}
Applying \eqref{pck-prod} and \eqref{hcoeff}, we obtain finally
\eq{
\sumi h_{i,l} \gamma_{i,j} &= M \prodk (-1)^{n_{j,k}n_{l,k}}\cdot \prodk \sqrt{\pck(n_{j\oplus l,k})}
\nonumber\\
&= M h_{j,l} \sqrt{P_{j\oplus l}}
}
which completes the proof of \eqref{G4}.
\end{IEEEproof}

\subsection{Two Consecutive Transforms}
Theorem~\ref{thD}, which characterizes the vectors obtained by applying the new transform and the HT sequentially to a given alphabet, was given in Section~\ref{Sec:preliminaries:new-transform}.
\begin{IEEEproof}[Proof of Theorem~\ref{thD}]
To prove \eqref{eq:thD1}, we first write $\tS$ as a function of $\mX$ using \eqref{HTdef} and \eqref{transform}, as illustrated in Fig.~\ref{visual_transform}. For $i=0,\ldots,M-1$,
\eq{
\tbs_i &= \frac{1}{M} \sumj \bs_j h_{i,j} \nonumber\\
&= \frac{1}{M} \sumj h_{i,j} \suml \bx_l \gamma_{j,l} \sqrt{P_l} \nonumber\\
&= \frac{1}{M} \suml \bx_l \sqrt{P_l} \sumj h_{i,j} \gamma_{j,l} \nonumber\\
&= \suml \bx_l h_{l,i} \sqrt{P_l P_{i \oplus l}}
}
using \eqref{G4} for the last equality. To obtain $\tS$ as a function of $\tX$, we express $\bx_l$ in terms of its inverse HT in \eqref{HTinv}, to obtain
\eqlab{D.1}{
\tbs_i &= \suml h_{l,i} \sqrt{P_l P_{i\oplus l}} \sumj \tbx_j h_{j,l} \nonumber\\
&= \sumj \tbx_j \eta_{i,j}
}
where
\eq{
\eta_{i,j} \triangleq \suml h_{l,i} h_{j,l} \sqrt{P_l P_{i \oplus l}}
.}
The coefficient $\eta_{i,j}$ can be expressed using \eqref{hcoeff} and \eqref{nbc-oplus} as
\eqlab{D.2a}{
\eta_{i,j} &= \suml \prodk (-1)^{n_{l,k}(n_{i,k}+n_{j,k})} \nonumber\\
&\qquad\cdot \sqrt{\pck(n_{l,k}) \pck(n_{i,k}\oplus n_{l,k})} \\
&= \prodk \bigg[ \sqrt{\pck(0)\pck(n_{i,k})} \nonumber\\
&\qquad + (-1)^{n_{i,k}+n_{j,k}}\sqrt{\pck(1)\pck(\bar{n}_{i,k})} \bigg] \label{D.2}
}
where to pass from \eqref{D.2a} to \eqref{D.2} we used Lemma~\ref{lemmaH} with $f_{k,u}=(-1)^{u(n_{i,k}+n_{j,k})}\sqrt{\pck(u) \pck(n_{i,k}\oplus u)}$. Depending on the value of $n_{i,k}$, this product can be partitioned into two subproducts
\eqlab{D.2b}{
\eta_{i,j} &= \prod_{\substack{k=0\\n_{i,k}=0}}^{m-1} \left[ \pck(0) + (-1)^{n_{j,k}} \pck(1) \right] \nonumber\\
&\quad \cdot \prod_{\substack{k=0\\n_{i,k} = 1}}^{m-1} \left[ \sqrt{\pck(0)\pck(1)} + (-1)^{\bar{n}_{j,k}}\sqrt{\pck(1)\pck(0)} \right] \nonumber \!\\
&= \psi_i \prod_{\substack{k=0\\n_{i,k} = 0}}^{m-1} \left[ \pck(0) + (-1)^{n_{j,k}} \pck(1) \right] \nonumber\\
&\quad \cdot \prod_{\substack{k=0\\n_{i,k}=1}}^{m-1} \left[ \frac{1+(-1)^{\bar{n}_{j,k}}}{2} \right]
}
making use of $\psi_i$ defined in \eqref{psi}. Since the two factors in \eqref{D.2b} depend on $n_{j,k}$ as
\eq{
\pck(0) + (-1)^{n_{j,k}} \pck(1) &= \begin{cases}
1, & n_{j,k} = 0,\\
\pck(0)-\pck(1), & n_{j,k} = 1,
\end{cases} \nonumber\\
\frac{1+(-1)^{\bar{n}_{j,k}}}{2} &= \begin{cases}
0, & n_{j,k} = 0, \\
1, & n_{j,k} = 1,
\end{cases}
}
they can be expanded into four factors as
\eqlab{D.2c}{
\eta_{i,j} &= \psi_i
\prod_{\substack{k=0\\n_{i,k} = 0\\n_{j,k} = 0}}^{m-1} 1
\cdot \prod_{\substack{k=0\\n_{i,k} = 0\\n_{j,k} = 1}}^{m-1} (\pck(0)-\pck(1)) \nonumber\\
&\qquad \cdot \prod_{\substack{k=0\\n_{i,k} = 1\\n_{j,k} = 0}}^{m-1} 0
\cdot \prod_{\substack{k=0\\n_{i,k} = 1\\n_{j,k} = 1}}^{m-1} 1
.}
The two products of ones can be omitted. The product of zeros may look strange but is perfectly legitimate; its value is by definition 1 if the set $\{k: [n_{i,k},n_{j,k}] = [1,0] \}$ is empty and 0 otherwise. Furthermore, since $\pck(0) - \pck(n_{j,k}) = 0$ for all $k$ in this set, the second and third factors of \eqref{D.2c} can be merged into
\eqlab{D.3}{
\eta_{i,j} = \psi_i \prod_{\substack{k=0\\n_{i,k} \ne n_{j,k}}}^{m-1} \left( \pck(0)-\pck(n_{j,k}) \right)
}
which together with \eqref{D.1} completes the proof of \eqref{eq:thD1}.

To prove the reverse relationship \eqref{eq:thD2}, we first multiply both sides of \eqref{D.1} with $h_{i,M-1} \eta_{l,i}/(\psi_i \psi_l)$ and sum over $i$:
\eqlab{D.4}{
\sumi \tbs_i \frac{h_{i,M-1}\eta_{l,i}}{\psi_i \psi_l} = \sumj \tbx_j\sumi \eta_{i,j} \frac{h_{i,M-1} \eta_{l,i}}{\psi_i\psi_l}
.}
The Hadamard coefficient $h_{i,M-1}$ can be factorized using \eqref{hcoeff} with $n_{M-1,k} = 1, \forall k$ as
\eqlab{hcoeff.M1}{
h_{i,M-1} = \prod_{k=0}^{m-1} (-1)^{n_{i,k}}
.}
Using \eqref{hcoeff.M1} in \eqref{D.4} and replacing the ratios $\eta_{i,j}/\psi_i$ and $\eta_{l,i}/\psi_l$ with their factorizations according to \eqref{D.3}, we obtain for the inner sum on the right-hand side of \eqref{D.4}
\eqlab{D.5}{
\sumi h_{i,M-1} \frac{\eta_{i,j}}{\psi_i}\frac{\eta_{l,i}}{\psi_l} = \sumi \prodk \phi_{k,n_{i,k}}
}
where
\eq{
\phi_{k,u} \triangleq \begin{cases}
(-1)^u, & u = n_{j,k} = n_{l,k}, \\
(-1)^u (\pck(0)-\pck(u)), & u = n_{j,k} \ne n_{l,k}, \\
(-1)^u (\pck(0)-\pck(n_{j,k})), & u = n_{l,k} \ne n_{j,k}, \\
(-1)^u (\pck(0)-\pck(n_{j,k})) (\pck(0)-\pck(u)), \hspace{-7em}\\ & u\ne n_{j,k} = n_{l,k}.
\end{cases}
}
The fourth case is always $0$, because either $\pck(0)-\pck(n_{j,k})$ or $\pck(0)-\pck(u)$ is $0$. Furthermore, the second and third cases can be combined into
\eqlab{D.6}{
\phi_{k,u} \triangleq \begin{cases}
(-1)^u, & u = n_{j,k} = n_{l,k}, \\
(-1)^u (\pck(0)-\pck(n_{j,k})), & n_{j,k} \ne n_{l,k}, \\
0, & u \ne n_{j,k} = n_{l,k}.
\end{cases}
}

We wish to apply Lemma~\ref{lemmaH} to the right-hand side of \eqref{D.5}. In order to do so, we first calculate from \eqref{D.6}
\eqlab{D.7}{
\phi_{k,0}+\phi_{k,1} &= \begin{cases}
1+0, & n_{j,k} = n_{l,k} = 0, \\
(\pck(0)-\pck(n_{j,k})) \hspace{-7em}\\ \quad -(\pck(0)-\pck(n_{j,k})),
	\hspace{-2em}&\hspace{2em} n_{j,k} \ne n_{l,k}, \\
0-1, & n_{j,k} = n_{l,k} = 1
\end{cases} \nonumber\\
&= \begin{cases}
(-1)^{n_{j,k}}, & n_{j,k} = n_{l,k}, \\
0, & n_{j,k} \ne n_{l,k}.
\end{cases}
}
Now by Lemma~\ref{lemmaH}, \eqref{D.5} can be expressed as
\eq{
\sumi h_{i,M-1} \frac{\eta_{i,j}}{\psi_i}\frac{\eta_{l,i}}{\psi_l}
= \prodk \left( \phi_{k,0}+\phi_{k,1} \right)
.}
By \eqref{D.7}, this product will be nonzero only if $j$ and $l$ match in \emph{all} bit positions $k=0,\ldots,m-1$, i.e., if $j=l$. Thus, again utilizing \eqref{hcoeff.M1},
\eqlab{D.8}{
\sumi h_{i,M-1} \frac{\eta_{i,j}}{\psi_i}\frac{\eta_{l,i}}{\psi_l}
&= \begin{cases}
\prodk (-1)^{n_{l,k}}, & j=l,\\
0, & j\ne l,
\end{cases} \nonumber\\
&= \begin{cases}
h_{l,M-1}, & j=l,\\
0, & j\ne l.
\end{cases}
}

Finally, we combine \eqref{D.4} and \eqref{D.8} into
\eqlab{D.9}{
\sumi \tbs_i \frac{h_{i,M-1} \eta_{l,i}}{\psi_i \psi_l} = h_{l,M-1} \tbx_l
.}
Dividing both sides by $h_{l,M-1}$ yields on the left-hand side the coefficent $h_{i,M-1}\eta_{l,i}/(h_{l,M-1}\psi_l)$, which can be expressed using \eqref{hcoeff} and \eqref{D.3} as
\eq{
\frac{h_{i,M-1} \eta_{l,i}}{h_{l,M-1} \psi_l} &= \prodk (-1)^{n_{i,k}-n_{l,k}} \nonumber\\
&\qquad\cdot \prod_{\substack{k=0\\n_{l,k}\ne n_{i,k}}}^ {m-1} \left( \pck(0)-\pck(n_{i,k}) \right) \nonumber\\
&= \prod_{\substack{k=0\\n_{i,k} \ne n_{l,k}}}^ {m-1} (-1)\left( \pck(0)-\pck(n_{i,k}) \right)
.}
This expression, substituted into \eqref{D.9}, completes the proof of \eqref{eq:thD2}.
\end{IEEEproof}

\balance %


\end{document}